\documentclass[graybox, secnum]{svmult}

\usepackage{mathptmx}       
\usepackage{helvet}         
\usepackage{courier}        
\usepackage{type1cm}        
\usepackage{makeidx}         
\usepackage{graphicx}        
\usepackage{multicol}        
\usepackage[bottom]{footmisc}
\usepackage{hyperref}        
\usepackage{soul}            
\hypersetup{colorlinks=true,urlcolor=blue}
\usepackage[square,numbers]{natbib}
\makeindex             

\begin{document}
\title*{Gamma-ray Polarimetry of Transient Sources with POLAR}
\author{Merlin Kole \thanks{corresponding author} and Jianchao Sun \thanks{corresponding author}}
\institute{Merlin Kole \at University of Geneva, DPNC, 24 Quai Ernest-Ansermet, CH-1211 Geneva, Switzerland \email{merlin.kole@unige.ch}
\and Jianchao Sun \at Key Laboratory of Particle Astrophysics, Institute of High Energy Physics, Chinese Academy of Sciences, Beijing 100049, China \email{sunjc@ihep.ac.cn}}
%
%
\maketitle
\abstract{
Polarization measurements of the gamma-ray component of transient sources are of great scientific interest, they are however, also highly challenging. This is due to the typical low signal to noise and the potential for significant systematic errors. Both issues are made worse by the transient nature of the events which prompt one to observe a large portion of the sky. The POLAR instrument was designed as a dedicated transient gamma-ray polarimeter. It made use of a large effective area and large field of view to maximize the signal to noise as well as the number of observed transients. Additionally, it was calibrated carefully on ground and in orbit to mitigate systematic errors. The main scientific goal of POLAR was to measure the polarization of the prompt emission of Gamma-Ray Bursts. During the 6 months operation in orbit POLAR observed 55 Gamma-Ray Bursts of which 14 were bright enough to allow for constraining polarization measurements. In this chapter we mainly discuss about the POLAR instrument along with the calibration and analysis procedures. Two analyses are described, the first is a straightforward method previously implemented in polarization measurements, whilst the second was developed to improve the sensitivity and to mitigate several of the issues with the former. Both methods are described in detail along with information on how these can be extended to perform time and energy resolved polarization measurements.}

\section{Keywords} 
Gamma-Ray, Transient, Gamma-Ray Burst, Polarization, POLAR, Compton Polarimetry
\section{Introduction}

The POLAR detector \cite{Produit2018} was developed by a Chinese, Swiss, Polish collaboration. Its scientific goal was to detect the polarization of the gamma-ray component of various transient sources such as Gamma-Ray Bursts (GRBs), Soft Gamma-Ray Repeaters (SGRs) and solar flares, with an emphasis on the first. GRBs are the brightest electromagnetic transients in the Universe, they have an extra-galactic origin and are among the most distant objects observed indicating the extreme amounts of energy involved. The duration of the gamma-ray emission, or prompt emission, from these phenomena has a bi-modal distribution \cite{GRB_duration}. The short population of GRBs, theorized to originate from compact object mergers, such as  neutron star - neutron star mergers or neutron star - black hole mergers, lasts from dozens of milliseconds up to 2 seconds. Long GRBs, theorized to be the result from the death of a massive star, last from approximately 2 seconds up to hundreds of seconds. Although much has been learnt during the last 50 years on these extremely energetic events, many questions on their nature remain unanswered. Examples regard topics such as the dominant emission mechanism responsible for the gamma-ray component, the emission location and the magnetic field structure in GRBs. 

Much of what we know from GRBs has been learnt from the abundance of spectral measurements, with Fermi-GBM alone having measured spectra of over 3000 GRBs \cite{Fermi_catalog}. However, a lack of clear, undebated, discoveries from spectral analyses in recent years indicates the need for additional parameters to be added to the study. A promising candidate capable of breaking the current stalemate in our understanding of GRBs is polarization measurements of the prompt GRB emission. Such measurements have been theorized to be a powerful probe for the emission mechanisms, geometry of the emission region as well as the magnetic field conditions present there. Additionally, the evolution of the polarization with time as well as its dependence on energy holds further information, prompting time and energy dependent studies. For a more detailed overview on the scientific importance of such measurements the reader is referred to works such as \cite{Covino_2016,Gill_2021}.

Although of great scientific interest, polarization measurements of GRB prompt emission at these energies have proven to be highly challenging \cite{McConnell_2016} and no measurements have thus far been precise enough to constrain theoretical models. This is in part due to the relatively low signal to noise for gamma-ray polarimeters, the potential for large systematic errors and finally the transient nature of GRBs. This transient nature requires a large Field of View (FoV) detector which further lowers the signal to noise. Finally, time and energy dependent polarization measurements further increase the difficulty due to a further reduction signal to noise. The POLAR detector was a fully dedicated gamma-ray transient polarimeter, as such, it was designed to minimize the above mentioned difficulties in polarimetry. The instrument observed half the sky continuously to maximize the number of observed GRBs, with a relatively large effective area to maximize the signal. Arguably most importantly, it was carefully calibrated before launch as well as in orbit to minimize the systematic errors. The instrument took data from the Tiangong-2 (TG-2) spacelab from September 2016 until April 2017 during which it observed a total of at least 55 GRBs, several solar flares, an SGR candidate, as well as a number of pulsars. In this chapter we will focus mainly on the analysis of the GRBs as most of the published results from POLAR focus on these objects. 

This chapter will start with a brief overview of the POLAR detector with an emphasis on its characteristics relevant for the polarization analysis. This is followed by a detailed discussion on the calibration of POLAR which was vital to minimize systematic errors and therefore to produce reliable measurements. Finally, the different analysis approaches will be described. First, the relatively straightforward analysis procedure applied for the first scientific publication will be presented along with its short comings. This is followed by a detailed description of the novel analysis method which allows for multi-instrument joint spectral and polarization measurements. This second analysis procedure was applied to both time integrated and time resolved analysis and is currently also being applied to energy resolved analysis, all of which will be discussed.

\section{Introduction to POLAR}

POLAR was a wide Field of View (FoV) gamma-ray detector sensitive in the 50-500 keV energy range which made use of a segmented plastic scintillator array. The instrument was fully dedicated to perform polarization measurements of transient sources, primarily GRBs. In this section we will outline the detection principle used in POLAR. This is followed by an overview of the detector itself where the main focus is on its features relevant to the analysis of polarization data. Finally, its performance characteristics will be presented in detail.

\subsection{Detection principle}

In the energy range of $\sim\,$10$\,$--$\,$1000$\;$keV the cross section for photo-absorption is low, especially in detector materials suitable for polarization measurements. This is due to the requirement for the secondary photon to travel a significant distance in the detector, which in turn requires gas based materials or those with low atomic numbers. These however, have a low stopping power for gamma-rays making them inefficient in this energy range. As pair production is not yet possible below 1 MeV, the only promising interaction method to employ for polarization measurements in this energy range is Compton scattering. The azimuthal Compton scattering angle of a photon interacting within a detector holds information on its polarization vector. This is illustrated below in the Klein-Nishina equation:

\begin{equation} \label{eq:1}
    \frac{d\sigma}{d\Omega} = \frac{r_o^2}{2}\frac{E'^2}{E^2}\left(\frac{E'}{E}+\frac{E}{E'}-2\sin^2\theta \cos^2\phi\right).
\end{equation}
where $r_0 = e^2/m_ec^2$ is the classical electron radius with $e$ being the elementary charge, $E$ is the initial photon energy, $E'$ is the final photon energy, $\theta$ is the polar scattering angle and $\phi$ is the azimuthal angle between the polarization vector of the incoming photon and the projection of the velocity vector of the final state electron on to the plane normal to the momentum vector of the photon. 

\begin{figure*}
    \centering
    \includegraphics[width=0.5\textwidth]{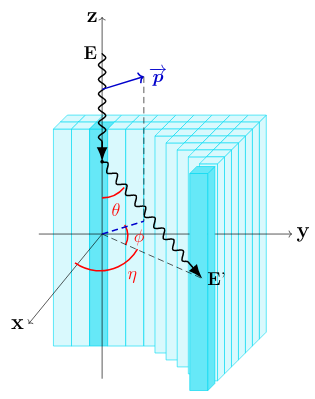}
    \caption{Schematic representation of a polarization measurement using a segmented detector. The incoming photon undergoes Compton scattering in one detector segment after which it undergoes a second interaction in a different segment. Its azimuthal Compton scattering angle in the detector $\eta$, which holds information on the polarization vector $\overrightarrow{p}$, can be calculated using the relative position of the two segments.}
    \label{fig:meas_principle}
\end{figure*}

Measuring the azimuthal Compton scattering angle requires two interactions of the photon within the polarimeter: an initial Compton scattering interaction followed by either a second Compton scattering or a photo-absorption. The key to a  polarization measurement is therefore the measurement of the two interaction positions within the detector. The position resolution for the two interaction locations determines the precision in the scattering angle measurement, which translates to the precision of the polarization measurement. Polarimeters therefore require either a position sensitive detector material, such as a silicon strip or pixel detector, or a segmented detector, for example a scintillator array. The former is complex and typically has a low cross section for Compton scattering in the $\sim\,$10$\,$--$\,$100$\;$keV energy range where the majority of the flux is. Scintillator arrays have a worse position resolution but allow for a  large, relatively low cost, low power, scaleable detector efficient at lower energies. An example of the polarization measurement principle employed with a segmented scintillator array is shown in figure \ref{fig:meas_principle}.

In figure \ref{fig:meas_principle} an incoming photon with energy $E$ enters the detector before Compton scattering in one detector segment. After continuing with energy $E'$ it interacts in a second segment. The angle between the two segments $\eta$ can be used to deduce the polarization vector of the incoming photon  $\overrightarrow{p}$. Due to the $180^\circ$ symmetry in the scattering angle distribution resulting from a polarized flux, identifying the order of the two interactions is not required. Simply detecting the two interactions within a small time window (to avoid chance coincidence events) therefore suffices for polarization measurements.

As illustrated in equation \ref{eq:1}, the ratio of photons scattering perpendicular to the polarization vector, over those scattering parallel to the polarization vector has both an energy dependence as well as a dependence on the polar scattering angle $\theta$. This ratio is largest when $E=E'$ which occurs in the Thompson regime, so when $E$ is small (several keV). Additionally, it is largest for $\theta$ around $90^\circ$ (the exact maximum depends on $E$). To maximize the sensitivity to polarization it is therefore preferable to design a detector where $\theta$ angles close to $90^\circ$. In figure \ref{fig:meas_principle} this would imply a large number of segments with a small height (along zenith direction). As the probability for an incoming photon to interact is inversely proportional to the height, a compromise has to be found between stopping power and polarization sensitivity. Additionally, a direct measurement of $\theta$ can also help to improve the sensitivity to polarization. Such a measurement requires a further segmentation along the zenith direction in figure \ref{fig:meas_principle} and can be complex to implement as it typically requires the addition of dead material and an increase in detector channels, which translates to a higher power consumption.

\subsection{The POLAR detector}

The POLAR detector makes use of a segmented scintillator array. The full plastic array comprised 1600 scintillator bars with a dimension of $5.8\times5.8\times176\,\mathrm{mm^3}$. The use of plastic scintillators over high Z materials, such as for example silicon detectors, increases the probability for photons to Compton scatter in the detector especially at energies below $100\,\mathrm{keV}$. The downside of a purely plastic based detector compared to a detector consisting of a combination of plastic and a high Z scintillator, such as that used in GAP \cite{Yonetoku}, is the lower stopping power after the initial interaction. After Compton scattering the photon can undergo several Compton scattering interactions or even leave the polarimeter, thereby reducing the efficiency. In the case of polarimeters like GAP the plastic detector is surrounded by high Z material increasing the probability for the photon to stop inside of the polarimeter. Such a design however, lowers the detector area, as photons entering the detector in the high Z material will likely be directly absorbed. Furthermore a full plastic array allows for a larger FoV as high Z materials placed on the periphery of the array will block incoming photons from the side from reaching the plastic scintillators. It should be noted here that the definition of the FoV of a gamma-ray polarimeter is somewhat complex and will be discussed in detail in section \ref{sec:characteristics}. A final downside of the POLAR design with only plastic scintillators is the low energy resolution. When incorporating high Z materials the energy resolution for the photo-absorption is relatively good. This energy measurement can be used to constrain the $\theta$ scattering angle, thereby allowing to improve the event selection in the analysis. As in POLAR the energy resolution is poor both for the scattering interaction and the photo-absorption, using the energy measurements to improve the polarization analysis is not possible, thereby reducing the sensitivity of the measurements.

\begin{figure*}
    \centering
    \includegraphics[width=0.95\textwidth]{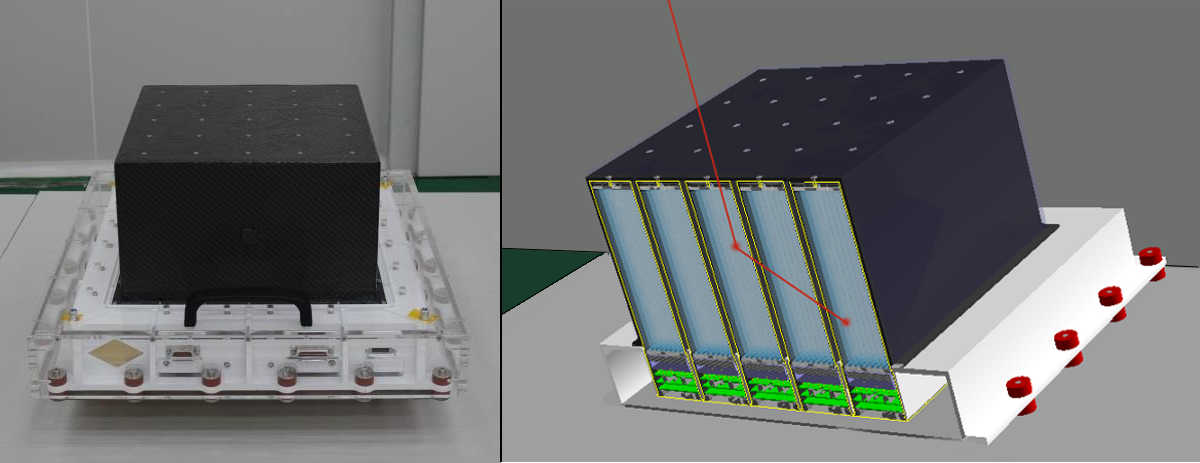}
    \caption{\textbf{Left:}The POLAR detector before launch. The carbon fibre top part can be seen while the white aluminium frame is covered with an acrylic glass cover to protect the fragile paint (this was removed before launch). \textbf{Right:} A schematic representation of the POLAR detector with a cut plane to show the inside. The scintillator bars, grouped in 25 modules, are shown in blue with below it the MAPMTs and the electronics. As an illustration of the detector effect an incoming photon is shown in red scattering in one scintillator and being absorbed in another.}
    \label{fig:polar_detector}
\end{figure*}

The plastic scintillator bars in POLAR are readout in groups of 64 by Multi-Anode Photo Multiplier Tubes (MAPMTs) each of which is readout using its own front-end electronics (FEE). The 64 bars, MAPMT and the FEE together form a detector module. The total of 25 detector modules communicate with the back-end electronics (BEE). The full detector, which is described in detail in \cite{Produit2018}, is shown in figure \ref{fig:polar_detector} along with a schematic indicating a typical event in the detector. 

The trigger logic of the instrument is handled by a combination of the FEE and BEE. In case at least 2 bars, which can be either in the same detector module or in different modules, have an energy deposition above a set threshold within the coincidence window the detector is readout. The coincidence window is 100 ns, which, given the typical trigger rate from a bright GRB being around 10 kHz means the probability for chance coincidence events can be ignored.

Events with a higher multiplicity (number of bars that trigger in the same coincidence window within one detector module) than 2 are also recorded, however, a maximum multiplicity of approximately 10 bars (this threshold was not sharp due to noise in the electronics) was set to minimize cosmic ray induced triggers. The $>2$ multiplicity events can be a result of incoming photons which scatter more than once in the detector, some of which can still be used in the analysis. They can however also be result of optical crosstalk in the detector. The MAPMTs used for scintillator readout in POLAR were the main cause of optical crosstalk. This effect had a significant influence on the data analysis. This crosstalk, which could be as high as $15-20\%$ from one scintillator to its neighbor was due to optical photons from one channel reaching the photo-cathode above a neighboring channel. This was possible due to the common entrance window of the MAPMT as well as an optical coupler placed between the MAPMT and the scintillators. The optical coupler, which ensured a good optical transmission from the scintillator into the MAPMT, also acted as a damper, protecting the fragile MAPMT window during launch. Due to the high level of optical crosstalk, energy depositions above $\sim100\,\mathrm{keV}$ in one scintillator bar could produce energy depositions in neighboring channels capable of triggering the electronics, thereby increasing the multiplicity and produce fake polarization events. Although it is possible to correct for this crosstalk in the analysis, as will be discussed in section \ref{sec:Data_processing}, it made it impossible to always distinguish a single energy deposition from 2 individual energy depositions in neighboring channels, thereby reducing the detection efficiency of the polarimeter.

\begin{figure*}
    \centering
    \includegraphics[width=0.75\textwidth]{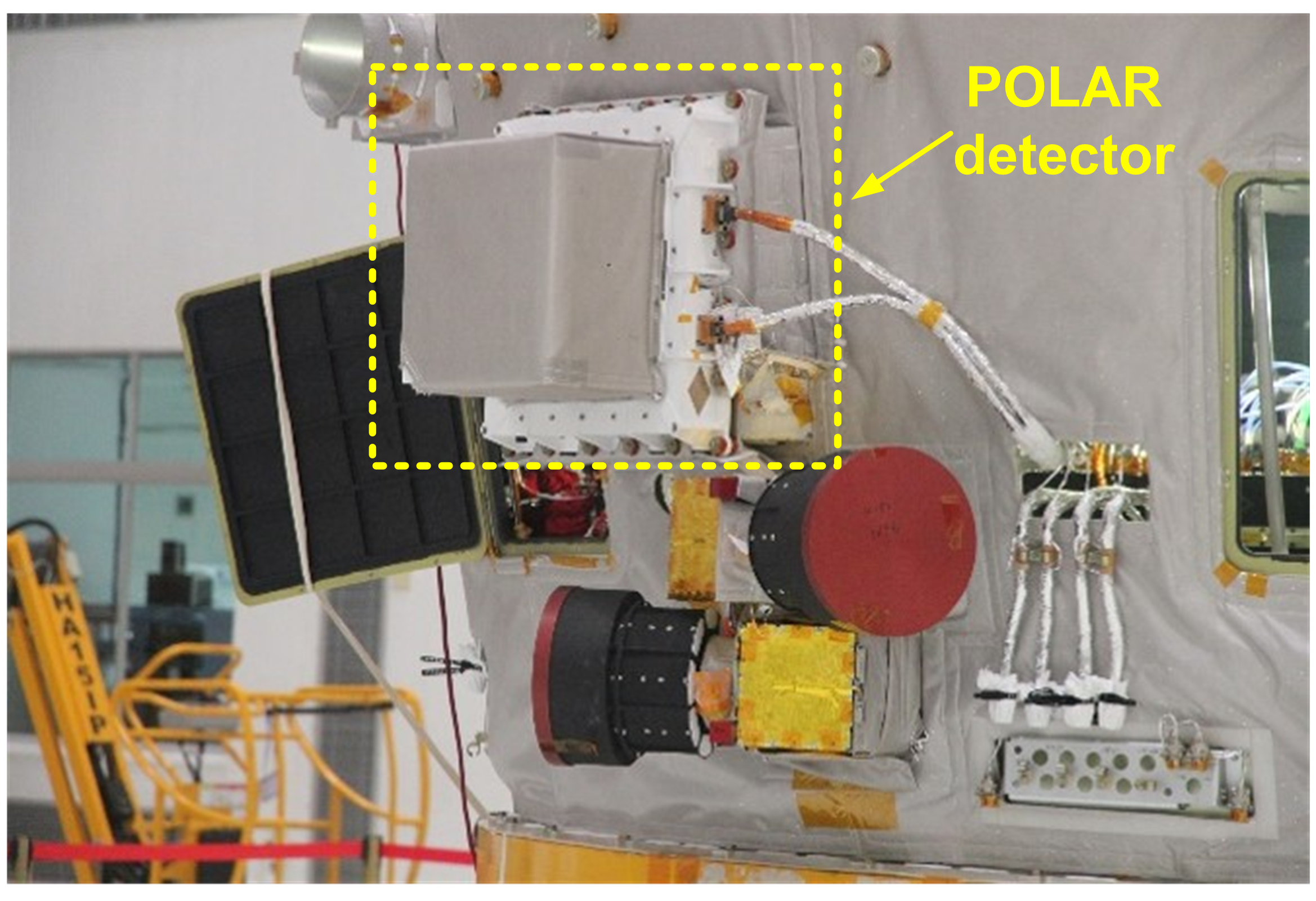}
    \caption{POLAR detector onboard TG-2. The photo shows where the detector is installed, while the interface electronics box is installed inside of TG-2 which is invisible in the photo.}
    \label{fig:POLAR_onboard_tg2}
\end{figure*}

\begin{figure*}
    \centering
    \includegraphics[width=0.75\textwidth]{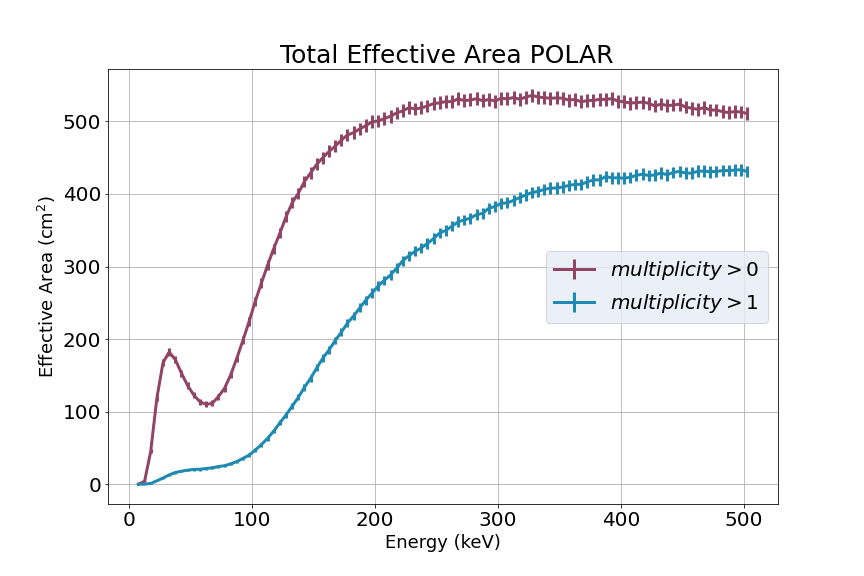}
    \caption{The total effective area of POLAR for a typical source location, here taken to be that of GRB 170114A which had an off-axis angle of $27.1^\circ$ with respect to the zenith of POLAR. The red line represents the effective area including all registered triggers and therefore includes single bar events which cannot be used for polarization. The blue lines contains all the stored triggers with a multiplicity larger than 1 which can possibly be used for the polarization analysis.}
    \label{fig:effectve_area_all}
\end{figure*}

The full POLAR detector was mounted on the TG-2 as shown in figure \ref{fig:POLAR_onboard_tg2}, the second Chinese spacelab, and was launched as part of the spacelab from the Jiuquan Satellite Launching Centre, China on September 15th 2016. The orbit and attitude, combined with the location of POLAR on the TG-2, ensured a continuous observation of half the sky without any Earth occultation throughout the mission. The placement on the TG-2 had several advantages, compared to a satellite based platform. The two major advantages relevant for data analysis are the relatively large power available for the payload and large data volume which could be produced and downloaded to Earth on a daily basis. These two advantages allowed POLAR to take data continuously throughout the mission without the need for changing to a special data taking mode in case of a GRB or other transient. As a result, the background was continuously monitored and could be studied in detail, greatly helping with the data analysis and the understanding of all relevant instrumental effects. Additionally, the data limit which was increased to 70 GB/day during the mission, allowed to not only keep events relevant for polarization measurements (those with at least 2 triggering bars), but also events with single energy depositions. Although not useful for polarization measurements directly, this data allowed to improve spectral measurements of GRBs. This improvement in spectral measurements, as will be discussed later in this chapter, reduces systematic errors on the polarization measurements. Furthermore, this additional data allowed to search for weak transient events as it provided POLAR with one of the largest effective areas of any gamma-ray detectors  in orbit at that time, as illustrated in figure \ref{fig:effectve_area_all}. The effective areas, both for all data and polarization data are shown in this figure. These effective areas are for a source with a location with respect to the POLAR zenith of $27.1^\circ$ which corresponds to that of GRB 170114A as observed by POLAR. Due to the cube like shape of the scintillator volume of POLAR the maximum effective area is for sources around $45^\circ$ off-axis, while the effective area for sources with an off-axis angle of $90^\circ$ is only slightly smaller than those observed on-axis.

A potential downside of the placement on the TG-2 was the possibility for material around the POLAR detector to affect the measurements as a result of photons scattering from neighboring materials. Apart from changing the measured flux and energy spectrum scattering will also alter the polarization of an incoming flux. This effect, which would also be present for satellite based detectors due to for example albedo flux from the atmosphere or solar panels, was studied in detail for POLAR as will be described in section \ref{sec:calibration}.


\subsection{Polarization sensitivity of POLAR} \label{sec:characteristics}

In this section we will discus the sensitivity of POLAR to polarization and the modulation curves of POLAR in detail. Modulation curve are the common term to refer to the histograms containing the azimuthal scattering angle distribution as measured by a detector. For a perfect detector the modulation curve is flat for an unpolarized flux while it shows a perfect sinusoidal function with a $180^\circ$ period for a $100\%$ polarized flux. The relative amplitude of the harmonic over its mean, as measured for a $100\%$ polarized flux, is referred to as the $M_{100}$. In an ideal polarization measurement a modulation curve is produced, the relative amplitude $M$ of the curve is measured and the polarization degree (PD) is retrieved using $\mathrm{PD}=\frac{M}{M_{100}}$, while the polarization angle (PA) can be deduced from the phase of the harmonic.

In practice however modulation curves are much more complex than often thought, especially for transient analysis, and understanding all its details is vital to perform proper data analysis. We will indicate the issues with basic data analysis together with shortcomings of some of the figures of merit used in polarimetry when performing data analysis with wide FoV detectors. 

\begin{figure*}
    \centering
    \includegraphics[width=0.75\textwidth]{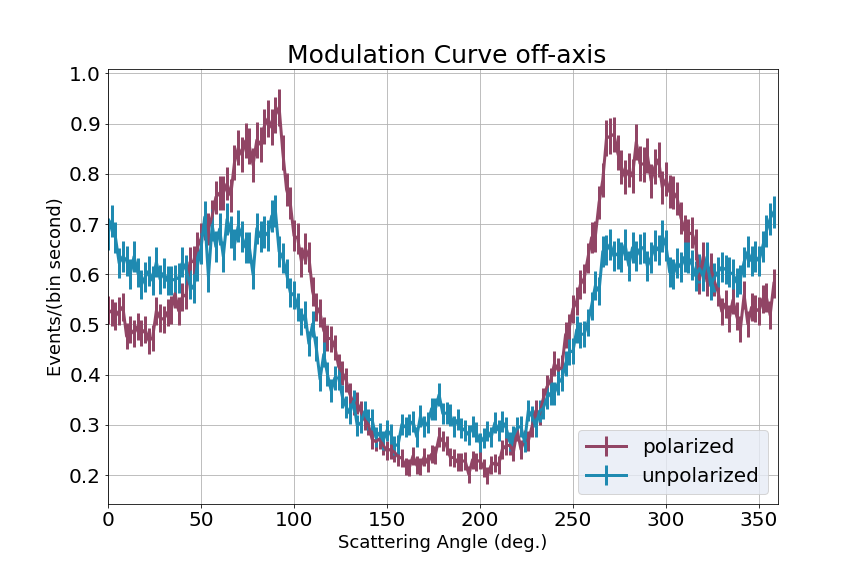}
    \caption{The simulated modulation curves for a GRB (170207A) with an off-axis angle of $70^\circ$ for a $100\%$ polarized flux (red) and an unpolarized one (blue). The scattering angle here is measured in the POLAR coordinate system.}
    \label{fig:typ_mod_curve}
\end{figure*}

First it is important to point out that a typical modulation curve obtained in a measurement of POLAR, or any wide FoV polarimeter, looks very different from a perfect $180^\circ$ modulation. This is due to instrumental effects as well as geometrical effects resulting from the incoming angle of the GRB with respect to POLAR. A typical modulation curve, obtained from the POLAR Monte Carlo (MC) software described in \cite{Kole_2017}, for an off-axis GRB with a $100\%$ polarized flux is shown (in red) along with that of an unpolarized flux (in blue) in figure \ref{fig:typ_mod_curve}. The measured modulation curve for this GRB (170207A), which looks similar to the unpolarized one in figure \ref{fig:typ_mod_curve}, can be found in \cite{Kole+20} along with the analysis results of this GRB.

Both the unpolarized and the polarized curve show a clear $360^\circ$ modulation with a minimum at $180^\circ$. This is a result of the GRB's large off-axis incoming angle of $70^\circ$. For large incoming angles the majority of photons interact in one of the sides of the detector. If they scatter from this location there is a larger probability for them to induce a second trigger when scattering into the detector, so along the momentum of the incoming photon. For photons scattering backwards, the majority will leave the polarimeter altogether leaving just one interaction, therefore not allowing for a scattering angle to be measured. For this GRB the incoming angle $\phi$ (as measured in the plane orthogonal to the zenith of the detector) was $-2^\circ$ with respect to the $0^\circ$ scattering angle axis in the POLAR coordinate system. As a result we see a maximum in the modulation curve at $0^\circ$ and a minimum at $180^\circ$. 

Apart from the clear $360^\circ$ component, small peaks every $90^\circ$ are visible in the modulation curve. These are a result of the square geometry of the POLAR scintillators. Due to this geometry, the distance to another bar is shorter when travelling perpendicular to the bar surfaces than when travelling diagonally, increasing the likelihood for photons to be detected in non diagonal bars. Additional, more smooth and less pronounced, effects are also present resulting from a variety of instrumental effects. Examples are differences in performance between different MAPMTs, differences between scintillator quality or even temperature gradients along the instrument affecting the electronics. Unlike in the case of point source observations, such as those performed by PoGOLite \cite{PoGOLite}, rotating the instrument along the observation axis during the observation in order to remove all such effects is not possible here. Although a rotation along the zenith axis would in theory remove instrument induced effects, the rotation would have to be too fast to be realistic. At least one rotation would have to be performed during each observation, which, for short GRB would require at least 10 revolutions per second. If on top of this one wants to perform time resolved studies of the GRB polarization this would have to be further increased. 

\begin{figure*}
    \centering
    \includegraphics[width=1\textwidth]{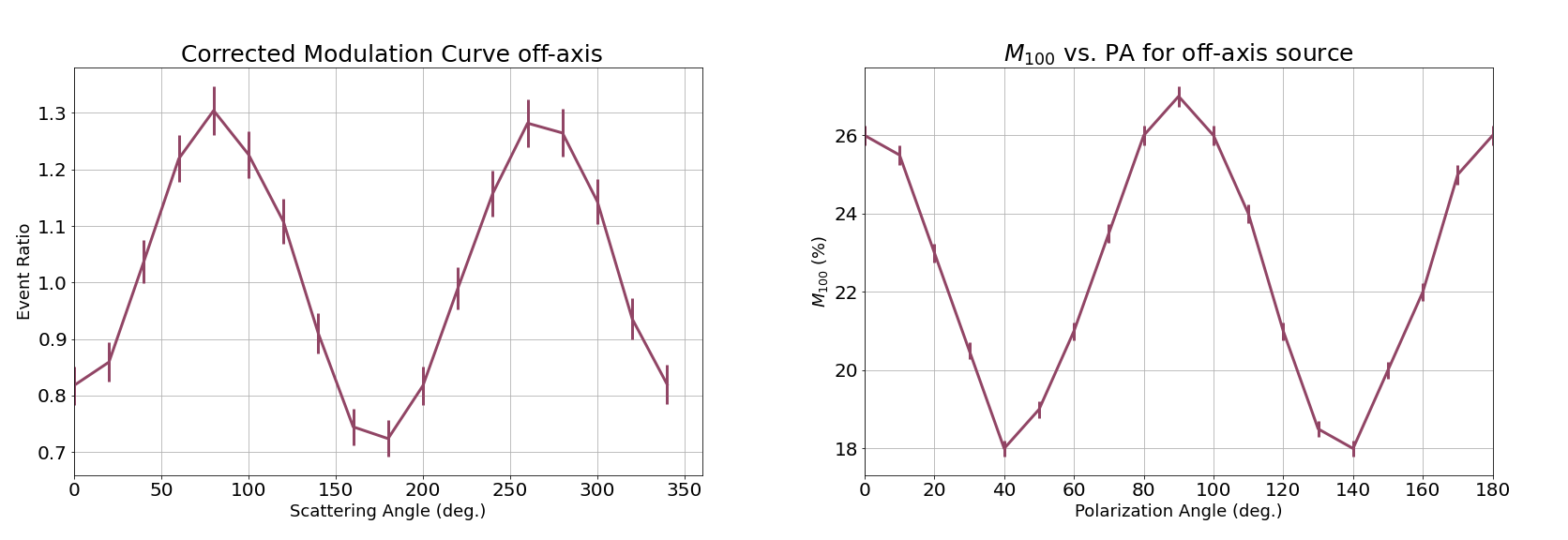}
    \caption{\textbf{Left:} The polarized curve shown in figure \ref{fig:typ_mod_curve} divided by the unpolarized one. The result is a corrected modulation curve (although no longer in unit of counts/(bin s)) where all the effects not induced by polarization are removed. \textbf{Right:} the simulated amplitude of the $180^\circ$ component for this GRB, or $M_{100}$ as a function of the polarization angle. It can be seen that, despite the large off-axis angle, the $M_{100}$ is still reasonably large when the photons are polarized along or perpendicular to the length of the scintillator bars. For polarization angles of $45^\circ$ along the bar axis the sensitivity to polarization drops.}
    \label{fig:mod_curves_off_axis}
\end{figure*}

On top of all of the features described above, a $180^\circ$ modulation is present in the red curve of figure \ref{fig:typ_mod_curve}. To make this clear one can divide the red curve by the blue one. By doing this one removes all the effects not induced by the polarization, as those are present in both curves. The result (after rebinning) can be seen on the left side of figure \ref{fig:mod_curves_off_axis}. The figure shows a clear $180^\circ$ modulation, as expected, however, it also shows a small $360^\circ$ modulation, which again is a result of the large off-axis incoming angle of this GRB. This remaining $360^\circ$ modulation is induced by photons scattering off the passive materials into the detector. For some polarization angles the photons will be more likely to scatter towards the detector than for others, additionally the polarization of those photons which have scattered on the passive material has changed during this interaction. The amplitude of this remaining $360^\circ$ modulation therefore depends on the intrinsic polarization angle. This effect indicates that even a modulation curve fully corrected for instrumental effects will not be a perfect $180^\circ$ harmonic and can therefore not simply be fitted as such. Although this effect is pronounced in the case of POLAR due to the influence of the large surface of the TG-2 on which it was mounted, this issue will exist for any wide FoV detector, due to the satellite structure. More importantly the Earth albedo can also induce such effects for instruments for which the Earth is visible, this Earth albedo component will be more complex than the relatively straightforward component from the TG-2 in POLAR.

\begin{figure*}
    \centering
    \includegraphics[width=1\textwidth]{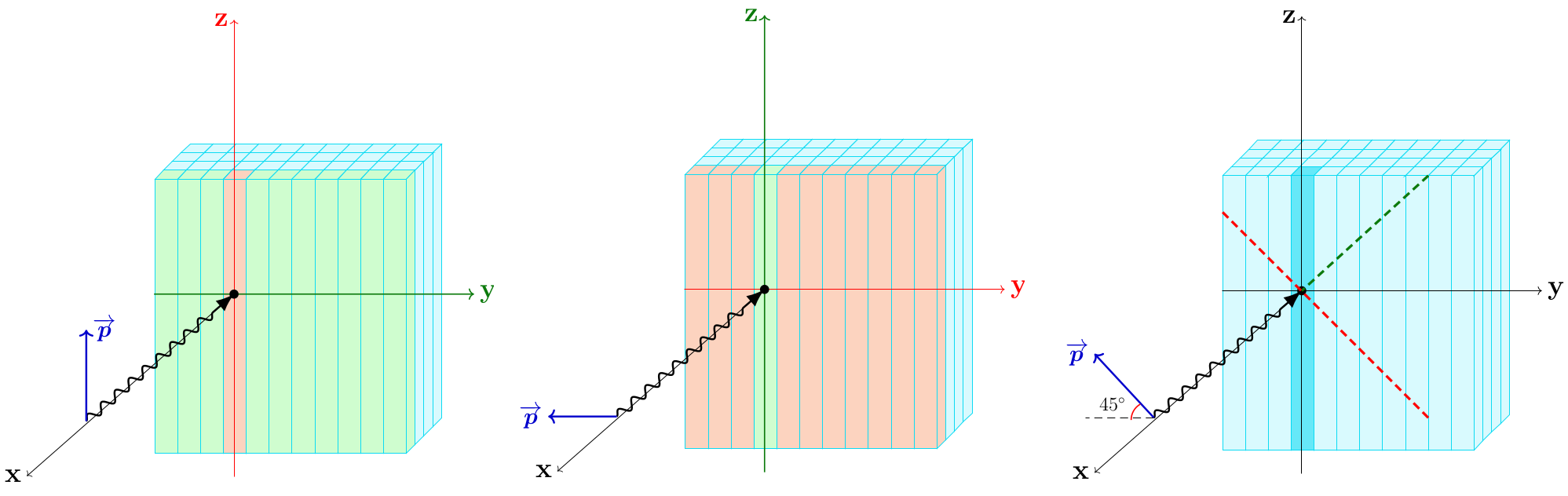}
    \caption{Schematic representation of the preferred scattering directions for photons entering the polarimeter from the side with three different polarization angles. \textbf{Left:} For a polarization along the z-axis the photons will have a larger likelihood to travel along the y-axis after scattering. The bars in green will therefore have a higher count rate than for a non-polarized beam while scattering twice in the same bar is less likely.\textbf{Middle:} For photons polarized along the y-axis more double interactions in the same bar are expected while interactions in the bars along the y-axis are decreased compared to a non-polarized beam. \textbf{Right:} For a polarization of $45^\circ$ with respect to the y-axis (or $135^\circ$ the favoured scattering direction is shown in green and the disfavoured in red. Although more interactions will appear in the top on one side and less in the bottom of the bars, the effect cancels out. As a result the modulation curve will look equal to the unpolarized one.}
    \label{fig:PA_dep}
\end{figure*}

A second more complex issue with the modulation curve is shown on the right side of figure \ref{fig:mod_curves_off_axis} which shows the amplitude of the $180^\circ$ component, which is the $M_{100}$, for different polarization angles. The figure shows a clear dependency which is again a result of the large off-axis angle in combination with the geometry of POLAR. To understand this effect, let's assume a GRB with an off-axis angle of $90^\circ$. The photons will enter POLAR at an angle perpendicular to the scintillator bar height as indicated in figure \ref{fig:PA_dep} where the incoming photon direction is set as the x-axis and the bar length the z-axis. For polarization along the z-axis, photons scattering with a polar angle ($\phi$) of $90^\circ$ in one bar will have a preferred azimuthal scattering angle $\theta$ towards neighboring bars. The result is an increased number of second interactions in the bars indicated in green. For photons polarized along the y-axis the preferred $\theta$ will be in the plane spanned by the bar height and the incoming direction, causing many photons to scatter a second time in the same bar, while the number of second interactions in the bars along the y-axis is now lower than for an unpolarized case. The modulation curves of both cases will look very different from one another and from the unpolarized case. Therefore, there is still a high polarization sensitivity for these angles, albeit lower than for an on-axis GRB (for which the $M_{100}$ is close to $40\%$). This is not the case for photon polarized at $45^\circ$ with respect to the z-axis. As illustrated on the right in figure \ref{fig:PA_dep}, the preferred scattering direction causes more photons to have a secondary interaction in the top of the bars in the positive y-direction, while interaction in the bottom of those bars are disfavoured. In the negative y-direction the situation is reversed. As there is no sensitivity along the z direction, the overall number of secondary interactions in any bar is equal to that of an unpolarized flux. As a result for such angles the instrument is fully insensitive to the polarization of the incoming photons. In case of a sensitivity along the z-axis this effect would be mitigated, however, using scintillator or semi-conductor bars such sensitivity is highly limited. As a result, this effect will be present in any wide FoV polarimeter without position sensitivity in 3 dimensions. Due to its geometry it is particularly clear in the POLAR data and becomes important for off-axis angles $>70^\circ$.

Not only this effect is important to understand in order to perform the data analysis correctly, but also it has an implication to the use of the $M_{100}$. Furthermore it has an influence on the often used figure of merit, the Minimal Detectable Polarization (MDP). The MPD represents the lowest level of polarization, for a given observation, which can be distinguished from being non-polarized with a set level of significance. For GRBs the MDP is best expressed as \cite{Weisskopf}:
\begin{equation}
    \mathrm{MDP} = \frac{2\sqrt{\mathrm{-ln}(1-C.L.)}}{M_{100}C_s}\sqrt{C_s+C_b}\,.
\end{equation}
Here $\mathrm{C.L.}$ is the confidence level, $C_s$ is the number of signal events and $C_b$ the number of background events.

As the $M_{100}$ and therefore the MDP has a dependency on the polarization angle, they can no longer be used directly to describe the sensitivity of a polarimeter to a certain source. In order to overcome this issue, one can take the averaged $M_{100}$ (and MDP) for all polarization angles to be able to continue to use these figures of merit for illustrative purposes. 

\section{The importance of calibration}

Arguably the main issue for gamma-ray polarization measurements, especially for the earliest ones, has been systematic errors. These systematic errors are generally a result of a lack of understanding of the sensitivity of the used detector to polarization. This in turn is often a result of insufficient on-ground calibration and the absence of calibration sources in space. An example of this issue are the early measurements performed using the RHESSI data for GRB 021206, the analysis of which resulted in 3 differing conclusions by 3 different teams \cite{Boggs, Rutledge, Wigger}. Although one could argue that this was a result of different event selections and potential mistakes in the analyses by some groups, the underlying reason is that the analysis methods applied could not be tested on reliable calibration data. If such data would have been available, the applied procedures could have been verified and optimized to produce a more conclusive result, even if, due to a limited signal to noise, this would not have produced a constraining measurement. 

RHESSI was not calibrated for polarization prior to launch as it was never intended as a polarimeter. However, also for dedicated polarimeters, systematic errors can easily dominate without a careful calibration of the detector. This is especially true for wide FoV detectors as the response of the detector depends not only on energy, but also on the location of the source on the sky with respect to the polarimeter. To further complicate issues, the performance of different parts of a detector will change with time, for example due to temperature variations. In the case of POLAR, temperature changes in the detector were within a few degrees \cite{Li_2018}, but remained non-negligible for the detector response. This dependency on the detector conditions, incoming angle and energy can only partly be measured directly on-ground, as it is not possible to measure all combinations at a calibration facility. The main objective of the on-ground calibration is therefore to verify MC simulations  capability to accurately produce the instrument response for every possible source and detector condition.

Due to the above mentioned reasons, the most challenging part of the data analysis for POLAR consisted of careful detector calibration, both on-ground and in-orbit. Subsequently this had to be translated into an accurate MC simulation framework. It was decided early on in the project that no analysis of scientific data would be performed until the required simulations were properly verified in order to avoid a repetition of issues found with previous GRB polarization measurements where different analyses found different results. In this section we will first illustrate the consequences of a poor instrument understanding on the final analysis result. This is followed by an overview of the different efforts made to mitigate such issues in the POLAR data analysis.

\subsection{Measuring zero polarization}

In practice it is impossible for a polarimeter to measure zero polarization. This is in part due to the requirement for an infinite amount of signal events, as can be seen in the MDP equation. Additionally, however, it is also due to the requirement to have a perfect understanding of the polarimeter. Here we illustrate this by looking at one of the most basic analysis methods applied to polarization data. The effects described here are however true, although more difficult to illustrate, for the more complex analysis methods described later in this chapter.

The typical analysis method consists of measuring the modulation curve of the source under study, subtracting potential background, and subsequently dividing the measured modulation curve by a simulated modulation curve for an unpolarized flux. For this simulated curve the flux has an energy spectrum and incoming angle equal to the observed source. Furthermore the detector conditions, such as temperature, should match those during the measurement. The division results in a corrected modulation curve, as illustrated in section \ref{sec:characteristics}. The corrected curve only contains the effects induced by the potential polarization from the observed flux while all other effects are divided out. This curve can be fitted with a harmonic function to retrieve PD and PA. Here, for simplicity, we ignore any potential $360^\circ$ component described in section \ref{sec:characteristics}. The relative amplitude of the harmonic function can be divided by the $M_{100}$ (the $M_{100}$ for the specific source including for example its energy spectrum) to acquire the polarization degree of the source.

\begin{figure*}
    \centering
    \includegraphics[width=1.0\textwidth]{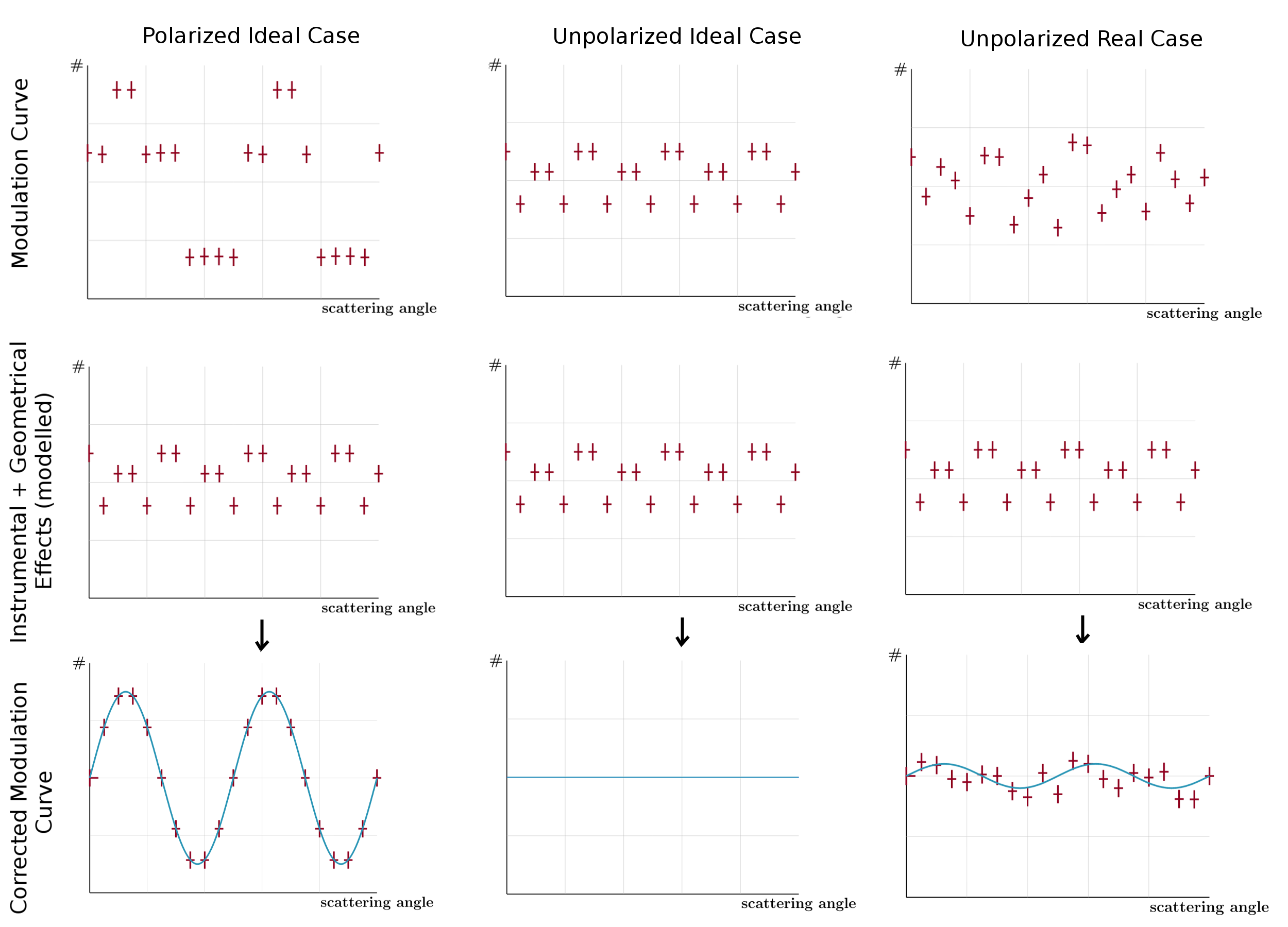}
    \caption{Illustration of recovering the polarization signal from a raw modulation curve. The left column illustrates 
    the ideal case with a high PD value, with a raw measured modulation curve (top), the perfectly simulated instrumental 
    and geometrical effects (middle) which pollute the raw modulation curve. The bottom-panel shows the modulation curve 
    after correction from the instrumental and geometrical effects which results in a perfect harmonic function. The middle 
    column illustrates the same but for an unpolarized signal resulting in a flat distribution. The right column shows the same for an unpolarized signal, however, 
    random small errors are added to represent instrumental effects which are not perfectly understood. The result is a non-flat distribution which, when fitted, shows a low level of polarization.}
    \label{fig:zero_polarization}
\end{figure*}

One clear issue with this method is that it will not always work due to the dependence of the $M_{100}$ on the polarization angle. Assuming that this is not an issue for the measurement here, and assuming a perfectly simulated modulation curve for the unpolarized source, this method allows you to measure zero polarization in case of an unpolarized source. This is illustrated in the middle column of figure \ref{fig:zero_polarization}. 

The understanding of the detector is, however, despite ones best efforts, never perfect and the curve one simulates for an unpolarized flux will not perfectly match the measured one. As a result, the corrected modulation curve, which in the case of an unpolarized flux should be completely flat, will always have some deviations from a flat line. As deviations in any direction will make the curve less flat, fitting it with a harmonic function will result in some amplitude. This will translate into a non-zero polarization, as shown in the right column of figure \ref{fig:zero_polarization}.

The above illustration serves to point out two important issues. First, any error in the simulated performance of the instrument, however small, will prevent you from measuring zero polarization, even if one has infinite signal events. Producing results where one measures a higher PD than the real one is therefore a more likely outcome of an analysis with significant errors in the simulation, calibration or analysis. Only for intrinsically highly polarized fluxes will such errors be able to produce lower measured polarization degrees than the true ones. 

The second point is that not only does one need to understand the detector as well as possible, but any remaining systematic effects also need to be known. These systematic induced effects in a non-polarized measurement indicate the lowest level of polarization one can measure with the detector. In case of POLAR, as will be described below, the limit was $\sim2\%$ despite the  significant efforts made over several years.

\subsection{On-ground calibration}\label{sec:calibration}

The polarization is measured indirectly using the anisotropy of the azimuthal scattering angles. This in turn is measured using coincident triggers of different detector channels. Similar to a polarization signal, a difference in the sensitivity of individual channels can cause an anisotropy in the scattering angle distribution. Having one detector channel with a higher sensitivity than all the others will lead to relatively more coincidences between that channel and all the others, which, will become a feature in the modulation curve. For this reason it does not suffice to just understand the overall response of a polarimeter but one needs to understand each individual channel in great detail.

\begin{figure*}
    \centering
    \includegraphics[width=0.8\textwidth]{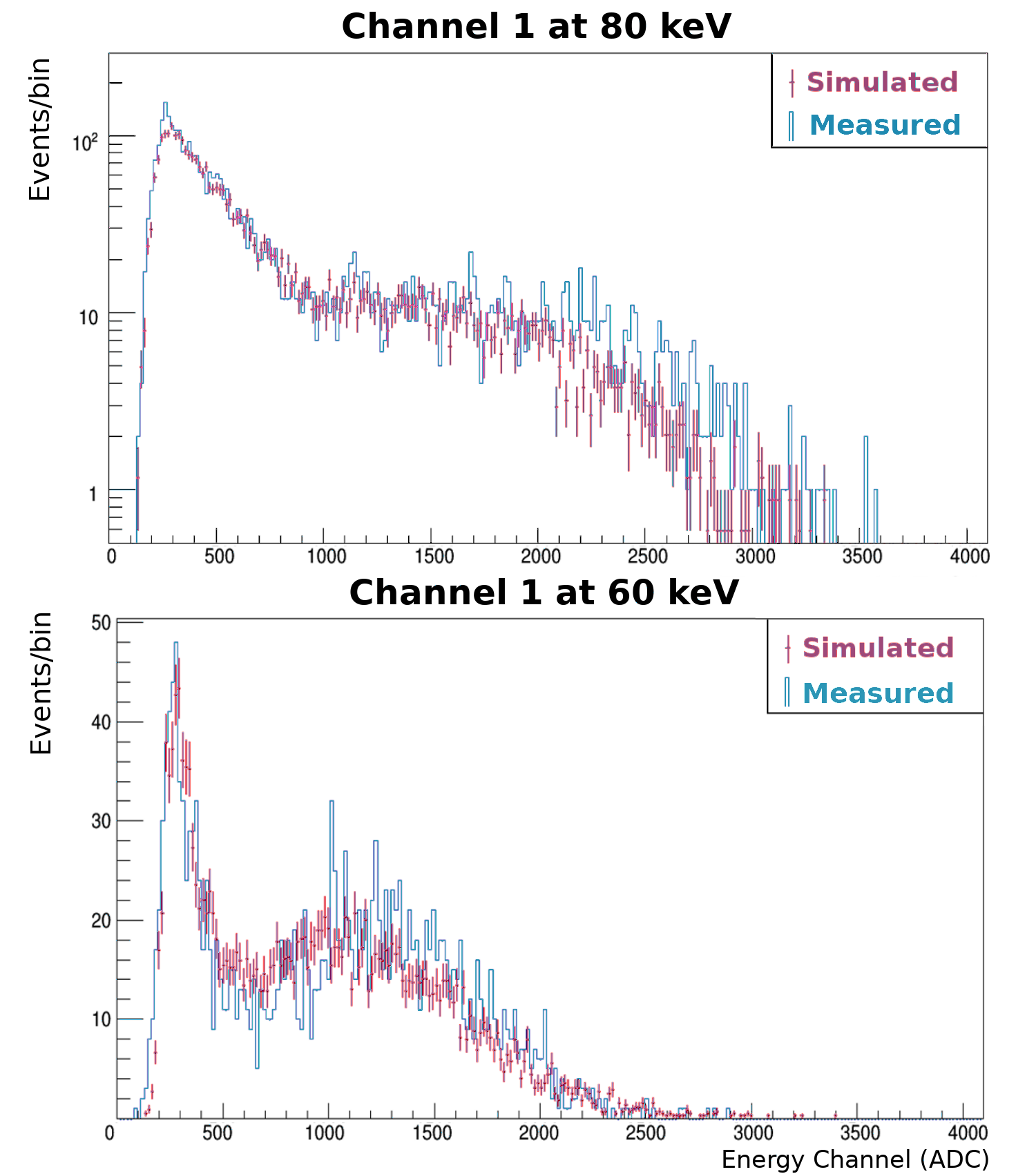}
    \caption{The spectra for one of the 1600 channels as measured while being irradiated with a 80 keV (top) and 60 keV (bottom) beam in blue. The spectra are compared with the simulation results (red) for this channel which can be seen to match well. It should be noted that the spectral shape for each of the 1600 channels looks different due to differences in gain, threshold, gain non-linearity and crosstalk.}
    \label{fig:channel_energy}
\end{figure*}

For this reason the on-ground calibration campaign of POLAR focused on measuring the noise, pedestal, gain, crosstalk, threshold and gain non-linearity of each channel including their dependence on temperature. Significant efforts were made to calibrate all these parameters using radioactive sources as well as synchrotron beams at the European Synchrotron Radiation Facility (ESRF) in Grenoble, France. Details on how all these parameters were measured for each of the 1600 channels are described in \cite{Kole_2017} and \cite{Li_2018}.  In parallel to the calibration of the 1600 channels an simulation framework was setup with the goal of accurately reproducing the instrument response. This framework included the physics simulation performed in Geant4 \cite{G4} which outputs the interaction location and (Birks' corrected) energy depositions for each interaction. These output parameters are then processed in a digitization software which transforms them into a data file equal to the digital output of the detector. This data can subsequently be processed using the analysis software as used on the real data. An effort was made to have the simulation output compatible to the real data as it allows to find any potential issues in the simulation as well as issues in the understanding of the real data. The details of this simulation framework are described in detail in \cite{Kole_2017} and \cite{Li_2018}. Finally the result of this effort was a simulation framework which could accurately reproduce the spectral response of each of the 1600 channels. An illustration of the performance can be seen in figure \ref{fig:channel_energy}, which shows the comparison of the simulated and real data for one specific detector channel to different beam energies. 

In order to study the instrument response to a polarized beam a calibration campaign was performed at ESRF with the POLAR flight model. During this campaign the instrument was irradiated with a near mono-energetic $100\%$ polarized beam (some higher harmonic energies were present in the beam). Irradiations were performed at 4 different energies (60, 80, 110 and 140 keV) and with 3 different incoming angles (on-axis, $30^\circ$ and $60^\circ$). Additionally, for each on-axis scan the instrument was irradiated with 2 different, perpendicular, polarization angles. As the beam only covers a small part of the full detector the detector surface was scanned for each configuration. During such a scan it was ensured that the centre of each channel was irradiated. It should be noted here that variations in the beam intensity during such a scan (which lasted up to 40 minutes) were recorded in order to apply them, together with the scanning pattern into the simulations.

\begin{figure*}
    \centering
    \includegraphics[width=0.8\textwidth]{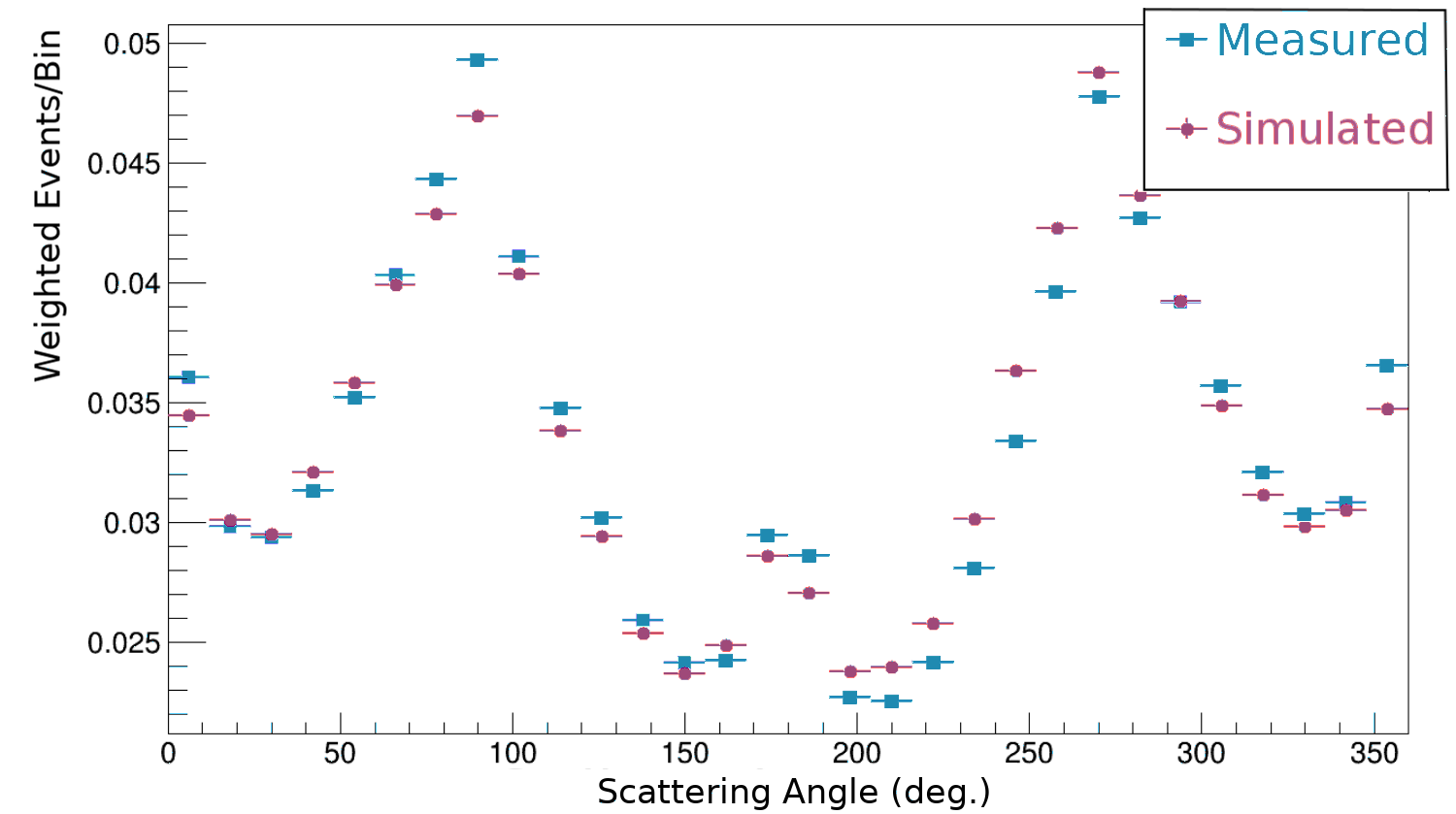}
    \caption{The measured modulation curve by POLAR from a 110 keV beam with an incoming angle of $60^\circ$ in blue. The simulated result is shown in red.}
    \label{fig:ESRF}
\end{figure*}

The details of the calibration campaign are summarized in \cite{Kole_2017} while an example of the measured and simulated modulation curves for one of the off-axis scans can be seen in figure \ref{fig:ESRF}. The modulation curves for each combination of energy and incoming angle were found to match the simulated ones within systematic errors as discussed in \cite{Kole_2017}. As the synchrotron facility cannot produce an unpolarized beam such a measurement was produced artificially by adding the measurements of two polarized measurements with perpendicular polarization angles. The final result, which is the closest to a perfectly unpolarized modulation curve which can be produced with this facility, resulted in a PD of $(1.6\pm0.1)\%$. This value was taken to be the minimal polarization which can be measured by POLAR, and as such used as a minimal systematic error throughout the analysis of the mission to account for imperfections in the simulated response. It should be noted that subsequent to this campaign several small issues in the simulations were solved which likely have improved the simulations and therefore reduced this systematic error. However, as these improvements were made during the flight of POLAR this could not be verified and the systematic error was kept for the remainder of the mission.

\subsection{In-orbit validation} \label{sec:inorbit_validation}

Subsequent to launch a detailed calibration campaign was performed in order to study potential changes in the instrument performance induced during the launch or installation of the payload. This was made possible thanks to 4 $^{22}\mathrm{Na}$ sources placed inside of the detector. These sources could, despite their low activity of approximately 100 Bq for each one, be used to measure the energy response of each of the channels to 511 keV photons. This is possible due to the back-to-back emission of the two 511 keV photons from $^{22}\mathrm{Na}$ which, combined with a precise knowledge on the location of the sources, allow for a dedicated event selection to search for emission from these calibration sources. Using days worth of data energy spectra from source photons can be produced which can be used to verify the spectral response of each channel as a function of temperature. Using this data the calibration parameters were verified and updated where needed, while further, previously not properly understood details, of the instrument response were found. These details, which stem from the electronics and required an update on how the non-linearity of the gain had to be simulated, are described in detail in \cite{Li_2018}, while the calibration method using the 511 keV photons is further detailed in \cite{Xiao_2018}. 

\begin{figure*}
    \centering
    \includegraphics[width=1.0\textwidth]{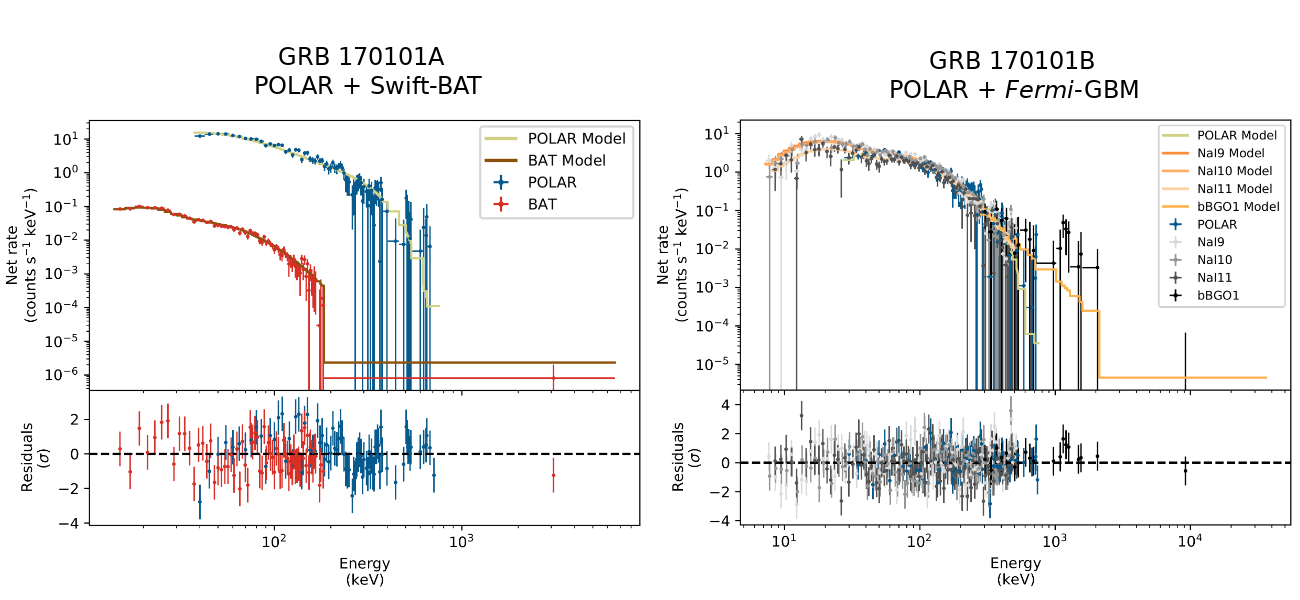}
    \caption{Examples of joint spectral fit results of GRBs observed by POLAR with Swift-BAT (left) and \textit{Fermi}-GBM (right). The number of counts for different (measured) energy bins in the different (sub-)detectors are shown (as data points) along with the modelled expected number of counts in that (sub-)detector for the best fitting spectrum (line). For  \textit{Fermi}-GBM 4 sub-detectors are shown along with POLAR, 3 of the detectors sensitive at lower energies (NaI) along with 1 detector sensitive at higher energies (BGO). The residuals for the different (sub-)detectors are shown in the bottom.}
    \label{fig:joint_fit}
\end{figure*}

In order to fully verify the understanding of the polarimeter, cross calibration studies were performed using the \texttt{3ML} \cite{3ML} framework. This framework, which was also used for the polarization analysis as described at the end of this chapter, allows to use data from different instrument to perform joint spectral analyses. In this case joint spectral fits were performed using data from GRBs jointly observed by POLAR and $\textit{Fermi}$-GBM and for 2 GRBs observed jointly by POLAR and Swift-BAT. In these joint spectral fits the official public instrument responses from $\textit{Fermi}$-GBM and Swift-BAT were used along with their public data. This is combined with the simulated response of POLAR for the specific GRB and the data from POLAR. The GRBs were fitted using a Band function \cite{Band} while a normalization constant to the absolute effective area was one of the free parameters. In figure \ref{fig:joint_fit} two of the joint fit results are shown as an example, the rest can be found in \cite{Kole+20}. The figure shows the best fitting spectrum as measured by both instruments in count space. This means it shows the expected number of detected counts per bin as measured in the respective detectors for the given spectrum. The residuals for the different detectors are shown in the bottom.

These results serve not only as a spectral fit, but also to study how well POLAR is calibrated against these two detectors. Figure \ref{fig:joint_fit} shows no significant outliers in the energy range, which is also the case for all other studied GRBs. Additionally, for all the studied GRBs, the normalization parameters for the overall effective area are found to be compatible to within $25\%$ of 1.0 when using POLAR and  \textit{Fermi}-GBM data, and below $10\%$ when using POLAR and \textit{Swift}-BAT. These values are similar to those found in catalog analyses of \textit{Fermi}-GBM for inter-calibration of the \textit{Fermi}-Large Area Telescope (LAT) and the different sub-detectors of GBM (the NaI normalization constants are around $5\%$, for the BGO the  normalization constants are of the order of $25\%$) \cite{catalogI}). This was the case for both on-axis GRBs as well as GRBs which were far off-axis. For these far off-axis GRBs materials surrounding POLAR play a role as photons scatter off from these areas into the detector. If not well represented in the simulations this can lead to systematic errors in both the spectral and polarization response. As no issues were seen in any of the GRBs studied it can therefore be concluded that the MC simulations are well capable of accurately reproducing the response of POLAR. 

It should finally be noted that this in-orbit calibration only verifies the simulated energy response and overall effective area. Cross calibrating with other polarimeters would allow to fully verify this response, assuming of course that the other polarimeter is well calibrated. The only gamma-ray instrument with polarization capabilities in operation during the POLAR mission was AstroSat CZTI \cite{Tanmoy+16}. Attempts to perform such a cross calibration have been proposed by the POLAR team but require the participation of the AstroSat CZTI team who's data is not public. It is anticipated that this activity will appear in the future after the internal calibration of the AstroSat CZTI instrument is finalized.

\section{$\chi^2$ analysis}\label{sec:Chi2}

Here we will present an overview of the first analysis performed with POLAR data which was conservatively performed using the $\chi^2$ analysis presented previously by the GAP collaboration \cite{GAP}. A more advanced method was developed afterwards which will be discussed in the next section.

\subsection{Data processing }\label{sec:Data_processing}

The general data processing pipeline of POLAR can be described as a series of corrections: subtraction of pedestal and common noise, data filtering, gain non-linearity correction, crosstalk correction and event energy reconstruction. For each step of the processing, the corresponding calibration parameters are required to perform such corrections which were acquired in flight during the commissioning phase as described in the previous section. The steps in the process, each of which is described in great detail in \cite{Li_2018}, are as follows:

\begin{itemize}
\item The pedestal is the signal caused by the leakage current of the electronics which was readout periodically with a frequency of 1 Hz. The common noise is the correlated signal shift of all the 64 channels of each detector modular unit during the readout procedure. As the pedestal and common noise are not part of the physical events, they should be removed. As the pedestal is monitored throughout the mission and the common noise is measured for each event subtracting these values is straightforward.

\item The POLAR data shows a non-negligible amount of abnormal events typically characterized by all the triggering channels having ADC values below their threshold level. These events often come in groups with a intermittent time equal exactly to the dead time required for recording the data, further indicating a non-physical origin. It was found that such events are induced by the effect of cosmic rays interacting in the front-end electronics and should thus be filtered out. This was performed using a cut value based on the event multiplicity and the maximum ADC value of the event. This cut had an efficiency of over $90\%$ while removing only a negligible fraction of the signal data. It should be noted that this step, as are all others, was also applied to simulated data in order to ensure it does not induce any effects in the polarization analysis.

\item Due to the readout logic of the 64 channels of the front-end electronics, the channel with the highest energy deposition within this modular unit initiates the waveform signal sampling of all channels within the detector modular. As there is a small dependency on the rise time of the signal on its total height this produces a small gain non-linearity effect where the ADC to keV of all channels in a modular unit depend on the highest ADC value measured in the unit. This effect, although minimal was measured, corrected for in the data, and implemented in the digitization of the simulation framework. 

\item As introduced in section \ref{sec:inorbit_validation}, four $^{22}\mathrm{Na}$ radioactive sources were installed inside the polarimeter, with activity of approximately 100 Bq for each source. The events of the two back-to-back emission 511 keV photons generated by the positrons released from the $^{22}\mathrm{Na}$ source were selected and used for energy calibration for POLAR in orbit. Using these values the gain of each channel could be calibrated. The source induced events in turn are easily identified and removed from the science analysis data. The same cut was applied to the simulated data where it removes a minimal fraction (below $0.1\%$) of the simulated signal events as it falsely identifies these as being source induced events.

\item As discussed earlier, the optical crosstalk mainly exists in the MAPMT window between adjacent channels, and has significant influence on the following data analysis. It should therefore be properly calibrated and corrected in the data. As the crosstalk in the data is measured in ADC and the ADC to keV conversion is different for each detector channel, and its gain is different, the correction of the crosstalk and the ADC to keV conversion have to be done together using a method described in \cite{Xiao_2018}. In this method the crosstalk and gain conversion for all 64 channels within a detector module are done together.
\end{itemize}

It should finally be noted that the temperature influence on all the calibration parameters, as well as the high voltage dependence of the detector gain, were measured and studied. Besides, the calibration parameters are also used in the digitization process of the Monte-Carlo simulations, which are a requisite for polarization analysis.

After the general data processing pipeline described above, the polarization analysis can be performed by producing one modulation curve from the measured data. In order to maximize the scientific potential of the POLAR data, studies were performed to optimize the MDP for typical GRBs in the event selection. This requires an optimization of the $M_{100}$, the signal to noise ratio and the overall number of signal events. It should be noted here that, as the MDP scales with the square root of the signal events and the signal to noise ratio, and linearly with the $M_{100}$ an increase in the latter is more valuable for the final result. Reducing the number of accepted signal events to improve the $M_{100}$ is therefore favourable. 

For this optimization all events with a multiplicity $>2$ were selected. Those with a multiplicity of 1 are not of interest for polarization studies. Subsequently, all events with an energy depositions outside of the dynamic range of the electronics were discarded as for such events the crosstalk could not be properly corrected for. It should be noted that this only accounts for $<1\%$ of the valid signal events. Finally, for each trigger event the two, non-adjacent, bars with the highest energy deposition were selected. If no 2 non-adjacent bars with energy depositions could be found, the event was not used in the polarization analysis. The removal of events consisting only of neighboring triggering bars was required as crosstalk induced triggers are capable of faking valid polarization events even after crosstalk correction. This is due to the correction using the mean crosstalk for the two channels in question, while the crosstalk varies significantly from event to event due to Poisson variations of the number of optical photons inducing the crosstalk. 

As valid polarization events with a short distance between the two interactions are not as sensitive to polarization as events with a larger distance between the two interactions, this cut greatly improves the MDP. Using this event selection the measured modulation curve was produced. This modulation curve is fitted to a simulated instrument response consisting of 6,060 simulated modulation curves to find the best-fitting polarization degree (PD) and polarization angle (PA) using the least $\chi^2$ method. More details about this method are discussed in the following subsections.

\subsection{GRB analysis}

The first sample of 5 GRBs analyzed by POLAR were selected based on the following criteria:

\begin{itemize}
 \item The GRB has been observed simultaneously by other instruments than POLAR and good measurement of both the spectrum and location can be provided.
 \item The fluence of the GRB, as provided by other instruments in the $10-1000\,\mathrm{keV}$ energy range, exceeds $5\times10^{-6}\mathrm{erg/cm^2}$.
 \item The incoming angle with respect to the POLAR instrument zenith, $\theta$, is below $45^\circ$.
\end{itemize}

All the selection criteria ensured a relatively high signal to noise and reduced as much as possible the systematic errors for the first GRB polarization analysis. This resulted in 6 GRBs, which are GRB 161129A, GRB 161218A, GRB 170101A, GRB 170114A, GRB 170127C and GRB 170206A. However, as the GRB 161129A happened on top of the tail of a large solar flare, it was not selected for polarization study to avoid any systematic effect coming from the solar flare events.

\begin{figure*}
    \centering
    \includegraphics[width=1.00\textwidth]{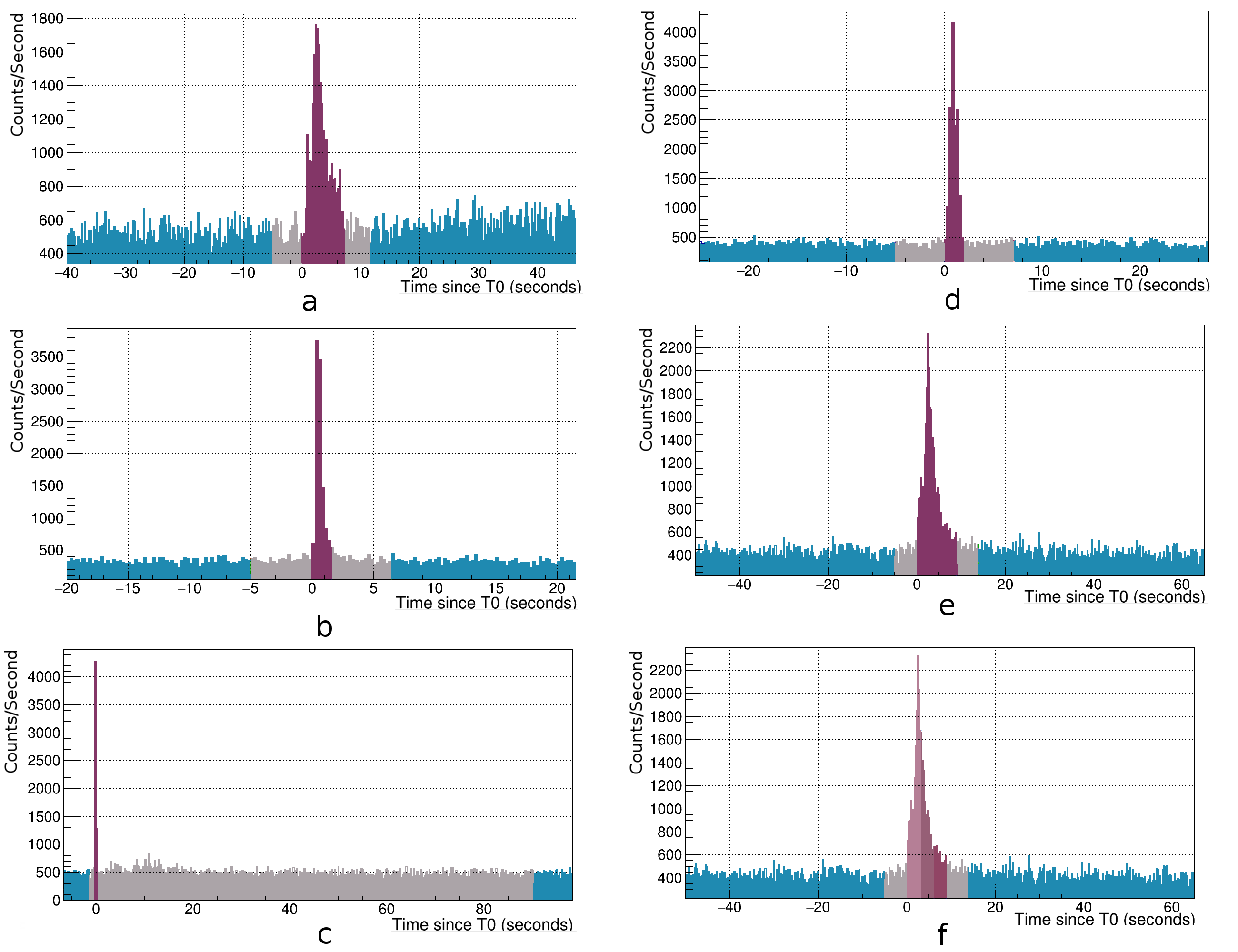}
    \caption{Light curves of all the first 5 studied GRBs sample. The selected signal region is shown in purple, the background region in blue and the non-used data in gray. Left from top to bottom are 161218A (a), 170101A (b) and 170127C (c); Right from top to bottom are 170206A (d), 170114A (e) and 170114A with additional details where the different time bins of 170114A, as used in time resolved studies, are indicated with different shades of purple (f).}
    \label{fig:lc}
\end{figure*}

For the measured modulation curve of each analyzed GRB the selected signal time intervals were based on the T90 values measured by POLAR. The background time interval was selected as two time intervals which are respectively before and after the GRB, as shown in figure \ref{fig:lc}. For GRB 170127C a longer period after the GRB was excluded from use as background as long duration, low intensity gamma-ray emission was seen in the POLAR data for several tens of seconds \cite{Kole+20}.

 The modulation curve of the background was subtracted from the modulation curve produced using data in the selected signal time interval. Here the relative difference between the live time of the signal and background intervals was taken into account, as well as the error propagation for each bin. As the background is relatively stable, the influence of the background selection on the final results was found to be negligible in a study described in detail in a dedicated section in the supplementary materials of \cite{Zhang2019}.

It should be noted that the dead time effect of the detector has also been taken into account for the modulation curve calculation, that is, the measured modulation curves were corrected based on all the events' counting rate and the dead time. For more details about the dead time correction, please refer to \cite{Wang}.

\subsection{Simulated response}

For each of the selected GRBs a 2d-grid in PD and PA of modulation curves was simulated. The simulations were performed using the framework previously introduced in section \ref{sec:calibration}. This framework consists of a Geant4 based physics simulation followed by the digitization which simulates the instrument electronics and trigger logic. 

In the physics simulation a detailed, well tested, mass model of the polarimeter is used together with a model of the TG-2 on which it was placed. As the influence of the TG-2 on the response could obviously not be performed on ground this remains a source of uncertainty. This is the main reasoning behind not including large off-axis GRBs in this study as for such GRBs the influence can become important. 

For each simulated GRB the best available spectral and location information was used. For each GRB spectral parameters, either from \textit{Fermi}-GBM, or from KONUS-Wind were available. The location of the GRB on the sky relative to POLAR was retrieved from location measurements of the specific GRB as reported by other instruments. At the relative location of the GRB with respect to POLAR a disk with a radius of 250 mm was placed from which a total of 5 million photons were emitted in the direction of POLAR. The size of the disk ensures a uniform irradiation of POLAR while also irradiating materials surrounding the polarimeter which could influence the final result. A total of 61 simulations were performed. One with an unpolarized flux and 60 with a PD of $100\%$ with varying PA (in steps of $3^\circ$).

The output from the physics simulations were subsequently processed in the digitization software to produce data equal to the measured data which enters the analysis pipeline described earlier in this section. This allows the simulated data to be processed in the same pipeline, thus ensuring the largest possible similarity between measured and simulated data. 

The final product is a set of 61 modulation curves, one for an unpolarized flux and 60 for a PD of $100\%$ for different PA values. The modulation curves for a PD between $0–100\%$ can be generated by mixing those with PD=0 and those with PD of $100\%$. This way, using the modulation curves of the 61
simulations, a total of 6,060 different simulated modulation curves was generated in the 2d plane of PA and PD with a step of $1\%$ in the PD direction and $3^\circ$ in the PA direction. 

\subsection{Systematic errors from spectral and localization}\label{sec:systematics}

The simulated modulation curves produced in the procedure described above are for a specific spectral shape and location. Differences in the spectral shape and location will result in a different modulation curves, meaning that uncertainties on the spectral and location parameters, which are non-negligible, carry over as uncertainties into the final polarization measurement. A second source of uncertainty in these simulated curves results from uncertainties on the in-orbit calibration parameters. 

In order to study these effects a set of 1000 simulations was performed for 1 specific GRB (170206A) where the spectral parameters, location parameters and calibration parameters were varied within their uncertainty distributions. A modulation curve was produced from each simulation to study the distribution of the bin contents of the modulation curves induced by the uncertainties in the GRB spectrum, location and calibration. After correcting for statistical fluctuations a relative variation of $4\%$ was found for the bins. This exercise was repeated for several PD and PA combinations each giving a similar result. Based on these findings a systematic error of $4\%$ was added to the data points in the simulated modulation curves to account for systematic errors.

\subsection{$\chi^2$ fitting}

In the final step of the analysis the measured modulation curve is fitted against the simulated modulation curves using the least $\chi^2$ method. The $\chi^2$ value for each simulated modulation curves is calculated using: 
\begin{equation}\label{equ:chi2}
\chi^2=\sum_{i=1}^n \frac{(X_i-Y_i)^2}{\varepsilon_i^2+\sigma_i^2},
\end{equation}
where $X_i$ and $Y_i$ are the counts of the bins of the measured and simulated modulation curves respectively, $\varepsilon_i^2$ and $\sigma_i^2$ are the uncertainties of $X_i$ and $Y_i$ respectively, and $n$ is the number of bins in the modulation curves. Both the measured and simulated modulation curves are normalized before calculating the $\chi^2$ value to ensure an equal total bin content. The uncertainties of the bins in the measured modulation curve are taken to be the statistical errors which are assumed to be normally distributed (see next section for a discussion on this assumption). For the simulated curves uncertainty is both the statistical one, which is negligible, and the systematic uncertainties discussed in section \ref{sec:systematics}.

\begin{figure*}
    \centering
    \includegraphics[width=1.00\textwidth]{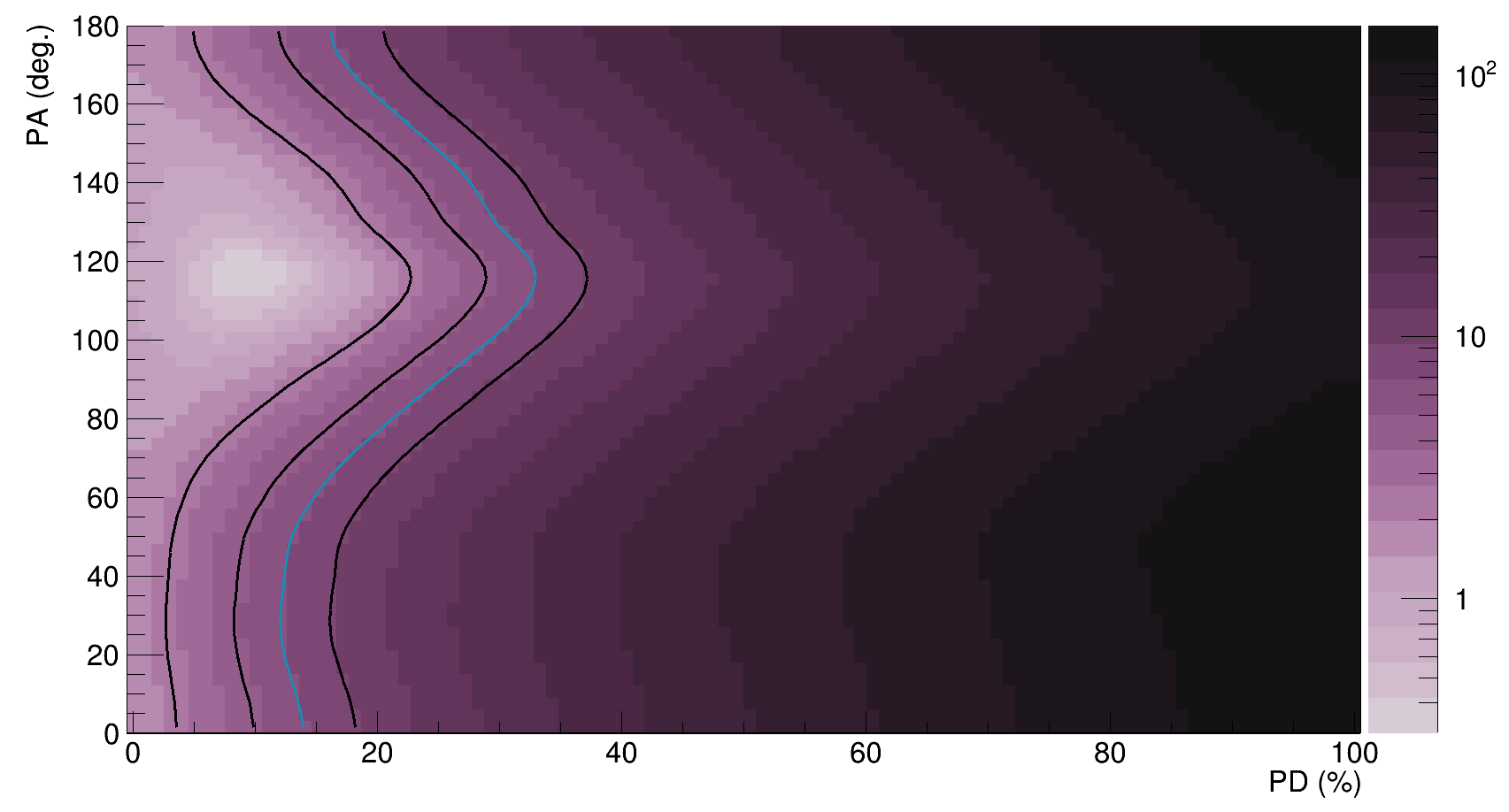}
    \caption{The $\Delta\chi^2$ map resulting from the analysis of GRB 170206A. The best fitting value can be seen for a PD of approximately $10\%$ with a PA of $117^\circ$. The 3 contours indicating the $1\sigma$, $2\sigma$ and $99\%$ confidence intervals in PD-PA space are shown in black, while that valid only in PD (so with 1 degree of freedom) is shown in blue. The maximum of this blue contour indicates the $99\%$ confidence upper limit in PD.}
    \label{fig:Delta_Chi2}
\end{figure*}

For each GRB an array of $60 \times 101$ of $\chi^2$ values corresponding to different PAs and PDs is then calculated. Finally, the PD and PA corresponding to the minimal $\chi^2$ which corresponds to the best fitting value, is compared in detail to the measured one to look for unexpected residuals which could indicate issues in the analysis. Non were found for the studied GRBs. The lowest $\chi^2$ is subtracted from all the values in the array to produce a $\Delta\chi^2$ map. An example of this for GRB 170206A is shown in figure \ref{fig:Delta_Chi2}. The confidence areas in the PA--PD space shown here were calculated using the value of $\Delta\chi^2$ as discussed in \cite{Avni1976}. The three black contours, from low to high PD, correspond to $\Delta\chi^2 = 2.28, 4.61$ and $9.21$. These values are the upper quantiles with probabilities of 32\%, 10\% and 1\% for the $\chi^2$ distribution with 2 degree of freedom. In order to calculate the upper (and potential lower) limit for PD alone the same is done but with only one degree of freedom for a probability of 1\%. This is shown with the blue contour which corresponds to the PD upper limit with a confidence level of $99\%$.

\subsection{Short comings of this method}

Although the method described in this section was verified to provide accurate results, as will be discussed in the next section, it has several shortcomings. The first regards the way the background is accounted for. In the above method, the background counts are simply subtracted (after correcting for the live time) from the signal counts. This can be done when assuming the number of counts are such that they follow a normal distribution, however, with tens of counts per bin this is not a completely correct assumption. As the data should still be handled as Poisson distributed, the subtraction leads to a loss of statistical information and wrong errors \cite{time_resolved}. Although not completely correct, this issue is minor, and subsequent studies with properly statistical handling of the data did not produce significantly different error distributions in the final result. This is however, possibly a result of the relative brightness of the 5 GRBs studied here. In order to be used on weaker GRBs, with less counts, a proper statistical handling of the data is required to ensure reliable results. 

A more significant issue with the data analysis is the propagation of systematic errors. In the above procedure, highly time consuming, simulations had to be performed in order to get an estimate of the influence of the errors of the spectral and location parameters on the final polarization result. While for the GRBs studied here all such errors were similar, one such simulation sessions sufficed. For a more complete analysis framework which can be applied to any GRB, direct propagation of the different errors to the polarization results would be of great benefit.

A third issue of the above method is the complete dependency on spectral and location information from other detectors. Whereas the spectral resolution and localization capabilities of POLAR are not as good as for example those of \textit{Fermi}-GBM, the data should be usable in case no spectral information from other detectors is available. For the second analysis, described in the next section, data from POLAR was used to produce both the spectrum, location and polarization for several GRBs. This was made possible by solving the above mentioned issue of error propagation as well as independent studies on the localization capabilities of POLAR \cite{Wang}.

The most important issue with the above mentioned method is not with the applied analysis methods, but rather with the lack of usability by those not familiar with the POLAR instrument and data. In order to meet the goal of producing a reliable set of polarization measurements it is important that such results can be reproduced by the wider community. This requires not only public data but also open source analysis tools which can be used by anyone who is interested and can be adapted to other instruments. 

Although the method described in this subsection can be published in an open source way, it is cumbersome to perform and requires a detailed knowledge of all the instrumental effects. Furthermore it requires the user to produce detailed MC simulation for each GRB to be studied. Therefore, in order to produce a fully reproducible set of results, the data as well as the instrument response should be made usable by a standard software framework. This allows for a reproduction of already published results and independent checks by the community.

\section{Bayesian time integrated analysis}

In order to improve on the short comings discussed before, a new method was developed based on forward folding. The analysis method was developed within the \texttt{3ML} framework \cite{3ML} such that it can be easily adapted for use by other polarimeters. Below we will discuss the method, how the analysis was expanded to allow to include data from other instruments and finally how more accurate results were produced both for time integrated and time resolved analysis. Finally we will touch upon the possibility to expand this method in the future for energy resolved analysis.

\subsection{Forward folding polarization data}

\begin{figure*}
    \centering
    \includegraphics[width=1.0\textwidth]{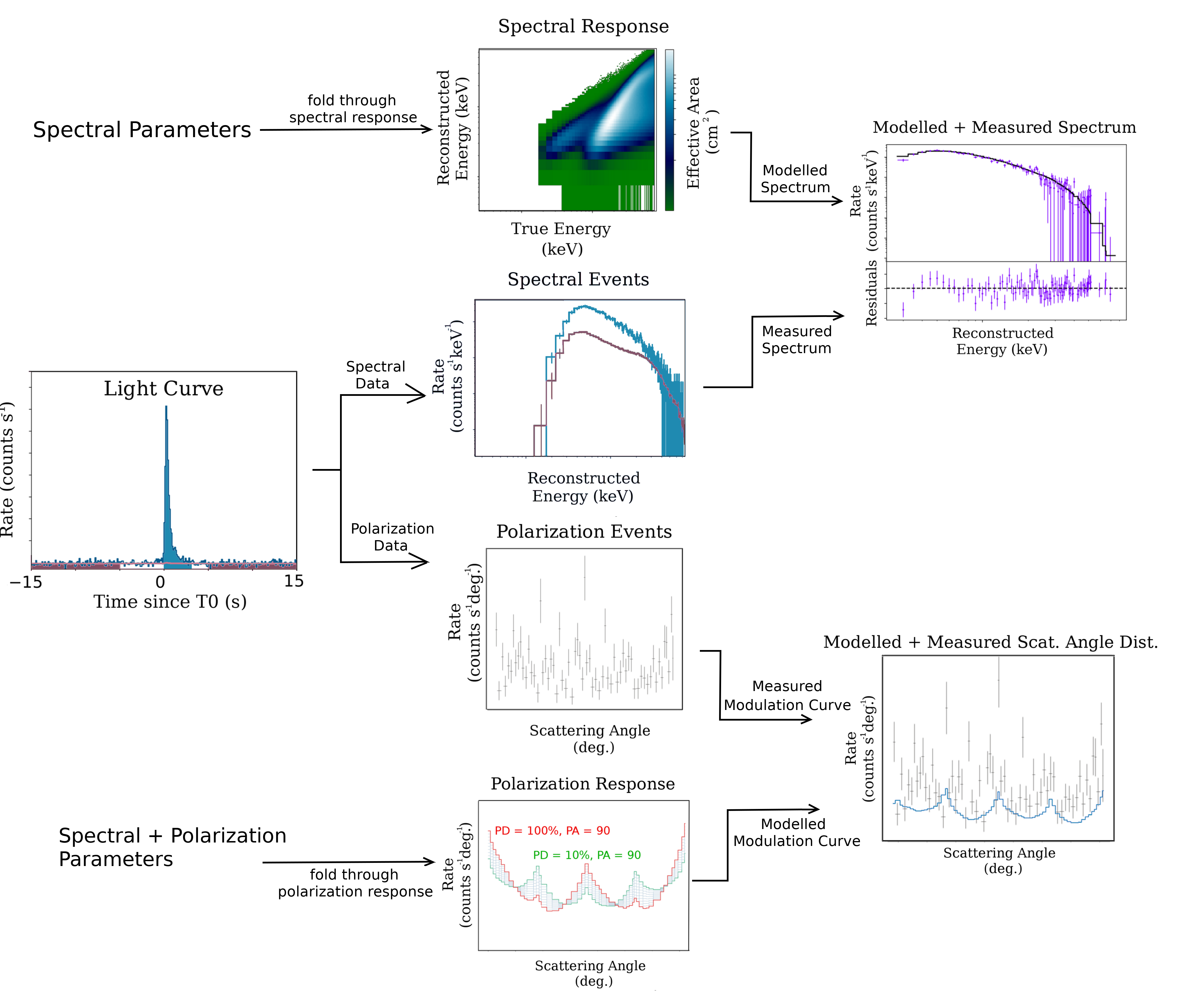}
    \caption{Schematic representation of the joint spectral polarization forward folding analysis. First the spectral and polarization data are selected from the light curve. A signal region is used to produce a spectrum (in count space) and a modulation curve. The background region is used to fit the background count rate as a function of time to produce the fitted background rate. A modelled energy spectrum is produced based on spectral parameters picked from the provided priors. This spectrum is folded through the energy response to produce a modelled spectrum (again in count space) which is used together with the measured spectrum to calculate a likelihood. In parallel, the same spectrum together with polarization parameters (also picked from a prior) are folded through the polarization response in order to produce a modelled modulation curve. Combined with the measured modulation curve a likelihood is also calculated here, which can be combined with the spectral likelihood. Repeating this process for different spectral and polarization parameters, a posterior distribution is produced. The spectral part shown here is performed in parallel using the data and responses for different detectors.}
    \label{fig:flow}
\end{figure*}

The developed polarization analysis method is based on the standard $\gamma$-ray forward-folding approach used for spectral fitting. In this method a detector's response ($R_{\gamma}$) for a given sky location ($\delta$) is produced. Typically this response consists of a 2d histogram containing the true photon energy ($\epsilon$) on one axis, and the energy measured by the specific detector (the reconstructed energy) on the other axis (an example is shown within figure \ref{fig:flow}. A proposed photon model ($n_{\gamma}$) is folded through the responses to produce detector count spectra ($n_{rec}$). Thus,

\begin{equation}
  \label{eq:2} n_{rec}^{i,j} = \int d\epsilon^j n_{\gamma}(d\epsilon,\bar{\psi}) R_{\gamma}^{i,j} \left(\delta \right)
\end{equation}
\noindent
where $i$ labels the detector and $j$ the reconstructed energy channel. Furthermore $d\epsilon$ is the true or latent photon energy and $\bar{\psi}$ are a set of photon model parameters (for example the parameters of the Band function) and
$\delta$ is the sky location of the GRB. Using the above equation one can calculate the number of counts in reconstructed energy bin $j$ for detector $i$  (so for example the bin of 15 keV in the energy spectrum as measured by POLAR)  for a given spectral model $n_{\gamma}$ using the detector response. Repeating this procedure for all reconstructed energy bins the result can be compared to the actual measured spectrum and subsequently repeated for different spectral parameters until, using some optimization algorithm, the modelled spectrum matches the measured one best. Finally, the same spectral model can be fitted to several detectors, in our case POLAR, \textit{Swift}-BAT and the several sub-detectors of \textit{Fermi}-GBM to optimize the spectral parameters using data from different instruments at the same time.

A similar method can be used for polarization analysis when one replaces the reconstructed energy bins by scattering angle bins from the measured modulation curve, and the detector response ($R_{\gamma}$) by a detector response which includes a polarization dependence. Meaning that for each true photon energy and polarization the response contains a reconstructed energy spectrum as well as a modulation curve. Mathematically one can write:

\begin{equation}
  \label{eq:1} n_{\theta}^{k} \left(\phi, \bar{p} \right) = \int d\epsilon^{j} n_{\gamma} \left(d\epsilon; \bar{\psi} \right) R_{\theta}^{j,k} \left(\epsilon, \phi, \bar{p} \right)
\end{equation}
where $n_{\theta}^{k}$ contains the number of counts in a scattering angle bin $k$, while $R_{\theta}^{j,k}$ is the response for scattering bin $k$. The response here is dependent on the energy, the polarization degree and the polarization angle. In practice for POLAR analysis, as used in \cite{time_resolved} and \cite{Kole+20}, both an energy and a polarization response was produced for a grid of true photon energies $\epsilon$, polarization degrees $\bar{p}$ and polarization angles $\phi$. For each combination of $\epsilon, \phi$ and $\bar{p}$ both an energy response in the form of a reconstructed energy distribution as well as a modulation curve was produced. This implies that similar to how one can produce a modelled energy spectra using the energy response and a spectral model, one can now produce a modelled spectrum as well as a modelled modulation curve for each possible combination of spectral model and polarization parameter. This modelled modulation curve can then be optimized with respect to the measured one in parallel to the optimization of the modelled spectrum with respect to the measured one. All in all, the method employed is the same here for both spectrum and polarization, however, the optimization is done for both spectral and polarization parameters at the same time, thereby naturally integrating the uncertainty of the fitted spectral parameters into the fitted polarization parameters.  The procedure is explained schematically in figure \ref{fig:flow}. 

The data was fitted to the model using a Bayesian probabilistic method already embedded within the 3ML framework, for details see \cite{time_resolved}. For this final step several valid methods are available, the Bayesian approach was chosen for this specific analysis because of a preference of the authors, it should however be noted that this final step should not affect the result significantly and that the main improvements over the original method came thanks to the forward folding technique. Additionally, the fitting was done using polarization parameters (picked from a flat prior) together with a synchrotron spectrum in the initial study presented in \cite{time_resolved}. When applying the procedure to the full POLAR catalog, however, the more model independent Band \cite{Band} function was used for the spectral fits. 

In the spectral part the spectral data from POLAR as well as that from either \textit{Fermi}-GBM or \textit{Swift} were used, while only data from POLAR was used in the polarization part. It should be noted that if data from other polarimeters were available this could easily be integrated in this analysis, similar to how data from different detectors is used in the spectral fit.

A second important thing to note here is that not only the spectral shape but also the overall intensity is folded through both the spectral and the polarization response. As a result the bins in the polarization response are in units of $\mathrm{counts}/(\mathrm{deg.}\,\mathrm{s})$ and the final modelled modulation curve includes also the expected number of counts per second. This modulation curve thus contains both information on the polarization and on the intensity of the spectrum which can be directly compared to the measured one, allowing to see any potential errors in the modelled sensitivity of the polarimeter. This significantly differs from the often used method where the integral of both the measured and simulated modulation curves are simply normalized to 1, thereby hiding potential errors in the modelled sensitivity.

Finally, it should be noted that the specific type of polarization response used here is not the only viable one when performing forward folding in polarization analysis. Instead of a modulation curve one can, for example, also use Stokes parameters, which would simplify the analysis as it replaces the large number of scattering angle bins of a modulation curve by 4 parameters. The reason Stokes parameters were not used here is the reduction in available information. In a modulation curve one sees all features, such as the instrumental ones discussed earlier in this chapter. If such effects are not well modelled this can easily be spotted when comparing the measured and the best fitting modelled modulation curve. When using Stokes parameters such problems would not be spotted so easily and potential systematic errors can therefore be missed in the analysis. The use of Stokes parameters in such analysis, as well as in other types of polarization analysis, are therefore appropriate only when the data and the model have been very carefully checked against one another and all potential sources of systematic errors are excluded or understood. It is therefore not excluded that Stokes based analysis will be performed with POLAR data in the future to simplify the analysis.

\subsection{Background modeling}

In the analysis described in section \ref{sec:Chi2} the background was originally subtracted from the signal. In this process a valid background interval was selected, a modulation curve produced for this time period and subtracted from the signal region using a scaling based on the relative time intervals. Two issues can be found with this method. First, the number of counts in the bins of the modulation curve are limited to several 10's of counts and therefore in the Poisson regime. Subtracting two Poisson distributions from one another results in a Skellam distribution which is not straightforward. To avoid complications in the statistical analysis the counts are often assumed to follow a normal distribution or the subtracted result is assumed to follow a Poisson distribution, but this can lead to improper handling of the errors especially when the counts are very low. For the POLAR analysis, where a normal distribution was assumed, this meant that a relatively low number of bins in the modulation curve should be used to avoid issues.

A second problem is that the subtracted background is taken as the average background during the selected interval. Any trends over time in the different scattering angle bins are ignored. This is not a problem when the background is very stable, however, this is often not the case on a satellite. 

Tests of the influence of different background periods and different lengths on the final polarization results were performed in \cite{Zhang2019}. These results showed no significant difference in the results, however, the tests were performed only on 5 GRBs which were particularly bright. To ensure a solid analysis framework usable on all GRBs in the POLAR catalog the method was changed to overcome the two issues.

Instead of taking an average background and subtracting it, the background rate, in all the scattering angle bins, was fitted using a polynomial. Using these fits the background during the signal was modelled and, instead of abstracting it from the data, included in the model. This way the pure data is fitted against a simulated signal plus a modelled background. This method avoids the potential issues with low counts, as no subtraction takes place, and therefore allows in principle for an infinite number of bins in the modulation curve. For the POLAR response a total of 360 scattering angle bins were produced. Apart from the statistical issues, the method also produces a more accurate representation of the background during the signal period as it includes potential trends present in different scattering angle bins.

\subsection{Adding data from other instruments}

An important feature of the framework developed in \texttt{3ML} is the possibility to add spectral data from other instruments to the analysis. This firstly allows to perform joint spectral and polarization fits. By using data from various spectrometers in the analysis the uncertainty on the spectrum, which caries over to the uncertainty on the polarization is reduced. Although POLAR has a poor energy resolution, this is partly offset by its large effective area allowing for some spectral measurement capabilities for broad band spectra (as opposed to for example line searches). Adding the data of POLAR to that of, for example, \textit{Fermi}-GBM, therefore can improve the spectral fit which will also improve the polarization fit. A second upside of the method which was pointed out before, is the natural inclusion of the spectral uncertainty into the polarization uncertainty.

The major upside of performing joint fits is however the possibility to perform cross calibration. By performing the joint spectral fits as shown in figure \ref{fig:joint_fit} it was assessed whether there are any significant systematic errors present in the POLAR response for any of the jointly observed GRBs. This study included off-axis GRBs such as 170207A and 170101B (on the right of figure \ref{fig:joint_fit}) non of which showed signs of systematics in either the overall effective area or the spectrum \cite{Kole+20}. This result indicates that, at least for the spectral part the simulations match the reality well, including for far off-axis GRBs. Although it cannot be checked in the same way for the polarization response, as no polarimeters with a verified calibration observed a large sample of the POLAR GRBs. However, as the spectral and polarization response were produced using the same software it is assumed that also the polarization response is reliable.

These results showed that GRBs with a large off-axis angle could now be analyzed. Additionally, for GRBs for which no public spectral data is available, the spectral data from POLAR was deemed reliable enough to perform a stand alone analysis where data from only POLAR was used to fit both the spectrum and the polarization. As a result of these checks on the reliability of the POLAR response, the number of GRBs usable for polarization analysis was increased from 5 to 14. 

It should finally be pointed out that the capability to include data from other instruments in the analysis can be expanded to polarization data as well. This implies that not only can data from a potential second polarimeter be used to improve the results, but more importantly cross calibration on the polarization side can be performed. A prime candidate for this is the data from AstroSat CZTI and POLAR for several GRBs observed by both instruments. A joint analysis would allow to understand the significantly different polarization levels reported by both collaborations, with AstroSat CZTI typically finding polarization degrees $>50\%$ \cite{Tanmoy+16} while the POLAR results favour polarization degrees $<50\%$ (see section \ref{sec:results}). Performing a joint analysis would allow to understand the potential sources of this discrepancy. 

\subsection{Time integrated Results}\label{sec:results}

Using the updated method described in this section the full POLAR catalog was analyzed. The selection criteria applied in the analysis discussed in section \ref{sec:Chi2} could be loosened thanks to the improvements. Firstly the criterion of the incoming angle was relaxed from only selecting GRBs within $45^\circ$ of the zenith of POLAR to selecting all within $90^\circ$, thereby increasing the sample by 7 GRBs. Additionally, GRBs for which no spectral information was available from other instruments were now included thanks to the capability to use the POLAR data for spectral fitting. Finally, the criterion on the brightness of the GRBs was relaxed due to the larger sensitivity of the improved method. Thanks to these second changes a further 2 GRBs could be added to the catalog, resulting in a total of 14 GRBs. 

\begin{figure*}
    \centering
    \includegraphics[width=0.80\textwidth]{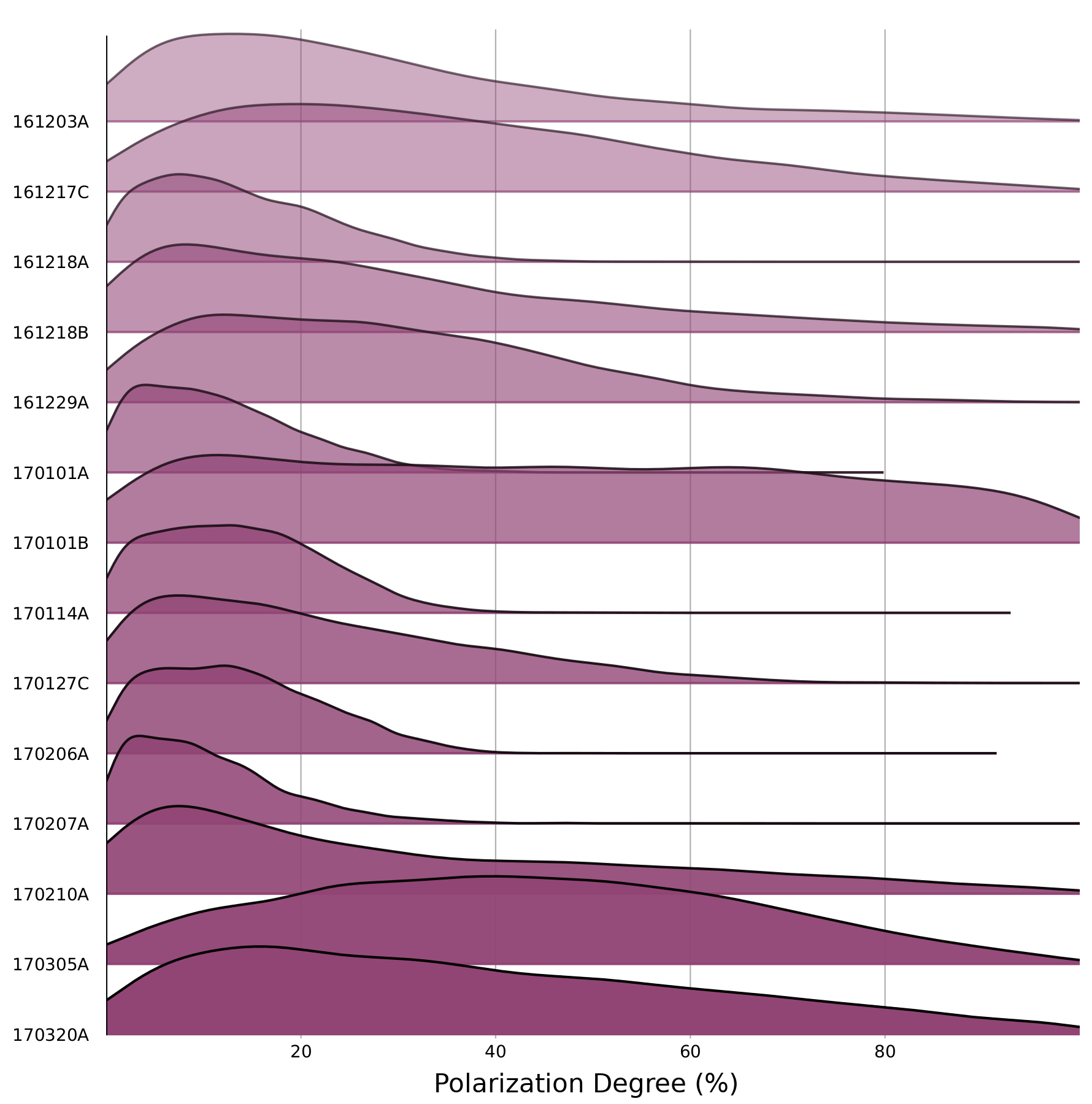}
    \caption{The results of the Bayesian forward folding analysis on the full POLAR catalog. The posterior distribution of the 14 individual GRBs is shown.}
    \label{fig:summary}
\end{figure*}

The posterior distributions of the polarization degree for these 14 GRBs can be seen in figure \ref{fig:summary}. For more details the reader is referred to \cite{Kole+20}. Firstly, the results for the 5 GRBs previously analyzed with the original method are fully compatible with those previously found. For these 5 GRBs a small reduction in the uncertainty is found, largely thanks to the reduction in systematic uncertainty due to the addition of spectral information. Additionally, for the 9 new GRBs the overall results are in agreement with the other 5. A low PD is favoured for all but one GRB which is 170101B, which has the largest uncertainty in PD. Overall it can be concluded that the results are fully compatible with an unpolarized flux, or a low PD, while polarization degrees above $50\%$ are disfavoured.

\subsection{Time resolved analysis}

Although the time integrated analysis results of POLAR appear to favour a low level of polarization, this does not imply that the emission is unpolarized. As adding two $100\%$ polarized fluxes with perpendicular polarization angles results in an unpolarized flux. The results discussed in the previous section could be the result of $100\%$ flux with an altering PA. In order to study this a time resolved analysis was performed where possible.

\begin{figure*}
    \centering
    \includegraphics[width=0.70\textwidth]{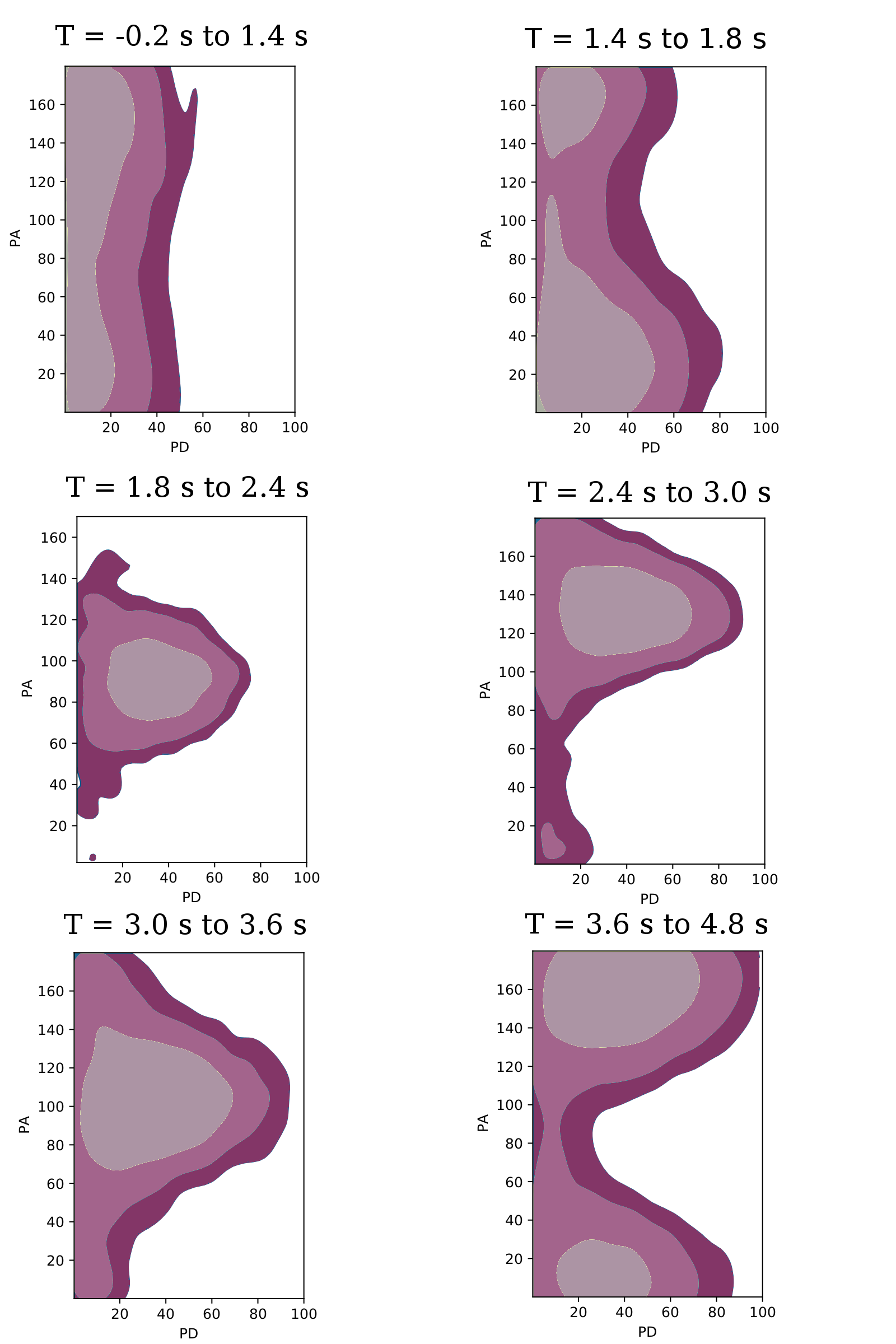}
    \caption{The posterior distribution of the PD and PA for the 6 time bins of GRB 170114A for which a constraining measurement is possible (for the other 3 the PD is fully unconstrained between 0 and $100\%$). The different colours indicate the $1\sigma$, $2\sigma$ and the $99\%$ confidence contours. It can be observed that while the time integrated result favours an unpolarized or very lowly flux, the time resolved study favours a PD around $30\%$ with a quickly changing PA.}
    \label{fig:time_res}
\end{figure*}

It is important to mention here that prior to performing the time resolved analysis a clear time bin selection strategy has to be agreed upon. Simply playing around with the time intervals and redoing the analysis many times is tempting and can easily lead to deceiving results with high levels of polarization. Although specifically important for time resolved analysis, it should also be noted that this is also valid for time integrated results. For example, different analyses of the same data of GRB 160821A resulted in different results \cite{Sharma+19, Tanmoy+16}, simply by selecting different time intervals. As  neither publication discusses the time interval selection or the reason for changing the selection between the two analyses, it makes it difficult to judge these results. In the POLAR analysis two methods were employed. For long multi-pulse GRBs, time intervals were selected to include single emission episodes. Where possible with multi-pulse GRBs or with bright single pulse GRBs, the pulse was divided into time intervals with an equal number of counts. A second, more sophisticated method, made use of the minimum variability timescale (MVT; see \cite{Vianello+18} for details). In this Bayesian blocks based method, the intervals are selected based on the minimum timescale above the Poisson noise floor during which variability exists in the data. 

For the GRBs for which the time integrated results already resulted in a large uncertainty in the polarization parameters performing a time resolved analysis is not useful. Therefore only the brightest GRBs from the 14 discussed previously resulted in constraining results. For all but 2 GRBs no signs of significant polarization, or therefore change in the polarization parameters throughout the GRB was found. This included for example GRBs such as 170207A which consists of 3 clearly separated emission episodes. 

The two GRBs for which a hint of polarization was found during the time resolved analysis were the two Fast Rising Exponential Decay (FRED) like pulses in the POLAR catalog, 170101A and 170114A. Although 170101A is not bright enough to perform really constraining measurements, 170114A was analyzed in detail using the MVT time interval selection as well as by dividing it simply in 3 equal intervals. In both cases a PD of around $30\%$ is found with a quickly changing PA. The results from the 6 time bins selected using the MVT method during which constraining measurements could be performed, are shown in figure \ref{fig:time_res}. Although the results are not constraining enough to exclude an unpolarized nature for all the time bins, they hint that while the overall GRB emission might be unpolarized, this could simply be an artefact of an evolution of the PA during a single emission episode. More detailed measurements are required to confirm this and to find the exact nature of the PA evolution.

\subsection{Energy resolved analysis}

Similar to how a time evolution of a polarization angle can mask the true polarization an evolution of the PD and PA with energy can mask the true polarization in an energy integrated analysis. Moreover, whereas theoretical predictions on the time evolution remain scarce, several recent works have predicted strong evolutions of the polarization degree with energy, see for example \cite{Ito, Lundman}. For these reasons performing energy dependent polarization measurements would be of significant interest. Such an analysis is however not straightforward for polarimeters as the true energy of the photons is not directly measured. For example, with POLAR one cannot simply produce a modulation curve for events originating from photons with a true energy in a specific range as the recorded energy cannot be directly translated to the true energy. While for other instruments, like for example the CZTI on AstroSat, the correlation between the true and the measured energy is more clear, the effect of energy dispersion will still play a role, making such a method not correct when performing energy dependent analysis.

The analysis framework described in this section can however be adapted to energy dependent polarization analysis. The instrument response produced for POLAR consists of a recorded energy spectrum and a modulation curve for each combination of the true energy (divided in bins of 5 keV ranging from 5 to 755 keV), the polarization degree (for PD of $0\%$ and $100\%$) and the polarization angle. For the time integrated analysis, the polarization and spectral parameters were folded through the response with the assumption that the PD and PA were independent on energy. However, by slightly modifying the method this assumption can be removed and different polarization models can be fitted to the data. An easy example, which would test the predictions from \cite{Ito}, is a joint spectral and polarization fit where the PD is 0 below an energy $E_{break}$ and a non-zero value above it. In this analysis $E_{break}$ could either be fixed, or set as a fitable parameter. This can be expanded to an analysis with several $E_{break}$ parameters between which the PD is independent, or even a full energy dependent PD profile. Of course, when making the model more complex the uncertainty on the final fit parameters will increase, setting a limit to the complexity which can be fitted to the existing POLAR data. Additionally, certain models can be tested more easily than others with the POLAR data. For example, the model presented in \cite{Lundman} predicts a high PD only at low energies (several tens of keV) where the sensitivity of POLAR is poor. Predictions made in \cite{Ito}, however, predict higher levels of PD above 100 keV, making it easier to test using the POLAR data. An effort to perform such analysis with POLAR data is ongoing and is foreseen to be finalized in early 2022. 

\section*{Acknowledgements}{
Merlin Kole acknowledges support from the Swiss National Science (SNF) foundation through the Ambizione program while Jianchao Sun acknowledges support from the National Natural Science Foundation of China (Grant No. 11961141013, 11503028), the Joint Research Fund in Astronomy under the cooperative agreement between the National Natural Science Foundation of China and the Chinese Academy of Sciences (Grant No. U1631242), the Xie Jialin Foundation of the Institute of High Energy Phsyics, Chinese Academy of Sciences (Grant No. 2019IHEPZZBS111), the National Basic Research Program (973 Program) of China (Grant No. 2014CB845800), and the Strategic Priority Research Program of the Chinese Academy of Sciences (Grant No. XDB23040400).
}

\vspace{0.3cm}


\end{document}